\documentclass[english]{article}
\usepackage[T1]{fontenc}
\usepackage[latin9]{inputenc}
\usepackage{geometry}
\usepackage{float}
\usepackage{url}
\usepackage{amsmath}
\usepackage{amssymb}
\usepackage{comment}
\usepackage{graphicx}
\usepackage{setspace}
\usepackage{color}
\usepackage{dsfont}
\usepackage[authoryear]{natbib}
\makeatletter

\providecommand{\tabularnewline}{\\}
\floatstyle{ruled}
\newfloat{algorithm}{tbp}{loa}
\providecommand{\algorithmname}{Algorithm}
\floatname{algorithm}{\protect\algorithmname}

\usepackage{algpseudocode}

\makeatother

\usepackage{babel}
\begin{document}
\title{Numerical Generalized Randomized HMC processes for restricted domains}
\author{Tore Selland Kleppe\footnote{Department of Mathematics and Physics, University of Stavanger, Stavanger, Norway. Email: tore.kleppe@uis.no}  and Roman Liesenfeld\footnote{Institute of Econometrics and Statistics, University of Cologne, Cologne, Germany}}

\maketitle

\begin{abstract}
We propose a generic approach for numerically efficient simulation from analytically intractable distributions with constrained support.
Our approach relies upon  Generalized Randomized Hamiltonian Monte Carlo (GRHMC) processes and combines these with a randomized  transition kernel that appropriately adjusts the Hamiltonian flow  at the boundary of the constrained domain, ensuring  that it remains within the domain.
The numerical implementation of this constrained GRHMC process exploits the sparsity of the randomized  transition kernel and the specific structure of the constraints so that the proposed approach  is numerically accurate, computationally fast  and operational even in high-dimensional applications.
We illustrate this approach  with posterior distributions of several Bayesian models with challenging  parameter domain constraints in applications to real-word data sets.
Building on the capability of GRHMC processes to efficiently explore otherwise challenging and high-dimensional posteriors, the proposed method expands the set of Bayesian models that can be analyzed by using the standard Markov-Chain Monte-Carlo (MCMC) methodology, 
As such, it can advance the development and use of Bayesian models with useful constrained priors, which are difficult to handle with existing methods.
The article is accompanied by an R-package (\url{https://github.com/torekleppe/pdmphmc}), which allows  for automatically implementing  GRHMC processes for arbitrary target distributions and domain constraints.
\end{abstract}

\section{Introduction}
In this paper, we use the  Generalized Randomized Hamiltonian Monte Carlo (GRHMC) methodolgy \citep{bou-rabee2017, kleppe_CTHMC} to develop a generic approach for efficient sampling from  analytically intractable continuous probability distributions with constrained support.
The need to simulate such constrained distributions is found in applications in many areas of statistics and econometrics. These range from applications of Bayesian Lasso approaches \citep{doi:10.1080/10618600.2013.788448} and analyses of limited dependent and binary variable models \citep{Albert:Chib:1993,Liesenfeld:Richard:Vogler:2017} to the use of Bayesian time series models with stationarity restrictions \citep{Koop:Potter:2011} and  Bayesian models with normalizing restrictions imposed to formally identify the model parameters, such as in neural network models \citep{9756596}.

The GRHMC approach modifies the standard HMC by randomizing the duration of the continuous-time Hamiltonian flow of the position for the  variables to be simulated and the momentum and allowing adaptive step sizes for the numerical integration used to approximate the Hamiltonian dynamics.
This special design makes the GRHMC more suitable for developing procedures for exploring constrained distributions than the standard HMC with its typically fixed step-size  numerical integration.

In a standard HMC implementation, if the Hamiltonian flow collides with the boundary of the restricted domain within the interval of a numerical integrator step, one could use an interpolation of the continuous-time Hamiltonian flow within that interval to accurately present the boundary collision event and then adjust this integrator step so that the Hamiltonian flow does not leave the constrained domain.
However, such an adjustment
would generally prevent the HMC integrator  from being time-reversible. 
As a result, the standard Metropolis-Hastings correction of the numerical integration is no longer applicable  because the numerically integrated trajectory  followed by
a momentum flip operation is no longer an involution mapping  \citep[see e.g.][Comment 2 after Theorem 2]{10.1214/aoap/1027961031}.

If, on the other hand, collisions are only allowed to occur at the integrator grid points, so that time reversibility is preserved,
then this can lead to a biased HMC approximation of the constrained target distribution, as these artificially shifted collisions deviate from the actual ones and therefore effectively occur outside the allowed domain  \citep[see e.g.][Figure 4]{10.1063/1.3573613}. This effect may be significant as Metropolis-Hastings corrected HMC methods often rely on rather coarse time steps. Further, it is not clear that trajectories that leave the allowed domain at some point will return to the allowed domain, which may lead to many rejected proposals.

An unbiased HMC procedure for such a case, but for a modified Hamiltonian  that leads to numerical integration steps that update  the position of the variables only parallel to the coordinate axes (and therefore is unlikely to scale well in higher dimensions), can be found in \citealp{10.1093/biomet/asz083}.

Within a GRHMC approach, these problems of the standard HMC can be circumvented, since in the numerical implementation of GRHMC processes,  approximation errors are not explicitly corrected but controlled by adaptive integrator step sizes,
which allows the  representation of the  Hamiltonian flow between integrator grid points  using appropriate numerical interpolations.
This means that  the collision events can be represented with arbitrarily high precision without having to forgo conventional Hamiltonian dynamics, which are known to scale well to high-dimensional target distributions \citep[see e.g.][]{10.1214/21-AIHP1197}.
Based on these insights, we develop a novel, generic and flexible GRHMC approach to explore numerically efficiently distributions with various types of constraints, which is also operational in high-dimensional applications.

Alternative approaches for simulation from constrained distributions include methods based on  Hamiltonian dynamics \citep[see e.g.][and references therein]{10.1063/1.3573613,doi:10.1080/10618600.2013.788448,10.1093/biomet/asz083} or on piecewise deterministic processes with  linear dynamics
\citep[see e.g.][]{BIERKENS2018148,2111.05859,2303.08023}. To ensure that the variables being simulated remain in their restricted   domain, these methods typically use  an   inelastic transition kernel inspired by deterministic physics to update the dynamics at the boundary of the restricted domain, which reflects the variables from the boundary back into the interior of the restricted domain.
However, as  illustrated in Section \ref{subsec:Boundary-kernels} below, such  a deterministic  inelastic boundary transition kernel can lead to repeated boundary collisions that occur with a very high frequency, which significantly  increases  the computational costs.
To limit such undesirable repeated boundary collision behavior, we propose a randomized  transition kernel for updating the Hamiltonian dynamics at the boundary. 
Since this proposed kernel is also sparsely parameterized, it also avoids suboptimal random walk-like dynamics in the simulated variables that are not directly affected by the constraints.
Both the randomness and the sparsity of the proposed boundary transition kernel contributes to the high numerical efficiency of our constrained GRHMC approach.

Another key element of our approach to improve its numerical efficiency by reducing computational cost is that it exploits the respective specific structure of the various constraints to identify when and where the (unconstrained) Hamiltonian flow reaches the boundary of the constrained domain.
For example, if the constraint has a linear form, then the time at which the numerical representation of the Hamilton flow hits the boundary  has an analytical closed form. This significantly reduces the computing time for identifying collision time and position compared to a general purpose numerical identification scheme.

A simple alternative strategy often used in derivative-based MCMC methods to take constraints into account is to use bijective transformations of the restricted variables, which lead to unconstrained target distributions \citep[see e.g.][Section 4.3]{JSSv076i01}.
However, for certain constraints, such a bijective variable transformation may not be readily available, and if it is, it may result in an unconstrained distribution with a complex non-linear dependence structure that hinders its fast exploration.
This is highlighted by simulation experiments in Section \ref{subsec:Unconstrained-vs-Transfromation}, where we analyse the relative merits of the proposed  constrained GRHMC method to this variable transformation approach.
It is found that even if there is a suitable variable transformation, it may still be advisable to simulate the original domain-constrained variables using the proposed method, even for simple models. 
Of course, in principle the relative performance of both approaches depends crucially on the type of curvature of the original and the transformed target distribution.

The constrained GRHMC methodology proposed in this paper  is implemented in the pdmphmc R-package \url{https://github.com/torekleppe /pdmphmc}.
It consists of a set of procedures which allows for easily and automatically implementing GRHMC processes for arbitrary target distributions with general domain constraints.
The code used to implement the models considered in this paper is available at  \url{https://github.com/torekleppe/constrainedPDMPHMC}.

The remainder of this paper is organized as follows. In Section 2 we outline GRHMC processes and their numerical implementation. Section 3 introduces GRHMC processes for simulating from constrained target distributions and Section 4 describes its numerical implementation for various types of constraints. There we also provide illustrative examples and results of simulation experiments that compare the constrained GRHMC with the variable transformation approach. Section 5 presents four empirical applications for real-word data sets. Section 6 concludes.

\section{Setup and background}
We aim to sample from  an analytically intractable continuous distribution with probability density function $\pi(\mathbf{q}),\;\mathbf{q}\in\mathbb{R}^{d}$,
and a density kernel (i.e. un-normalized density) $\tilde{\pi}(\mathbf{q})\propto\pi(\mathbf{q})$, which can be point-wise evaluated.
For numerical stability and efficiency of HMC procedures, it often proves to be advantageous not to sample $\pi$ directly, but rather a standardized version of the form $\bar\pi(\bar{\mathbf{q}})\propto\pi(\mathbf{m}+\mathbf{S}\bar{\mathbf{q}}),\;\bar{\mathbf{q}}\in\mathbb{R}^{d}$, where $\mathbf{m}$  represents the  location of $\pi$, and $\mathbf{S}$ is a diagonal matrix with  diagonal elements representing the marginal scales of each element in $\mathbf{q}$ under $\pi$.
(In the  applications presented below, for $\mathbf{m}$ and $\mathbf{S}$, we use estimates of the mean and marginal standard deviations of $\mathbf{q}$ under $\pi$, respectively, which arise in  the burn-in period of our proposed procedure.)
Obviously, with simulated values for $\bar{\mathbf{q}}$  targeting $\bar{\pi}$, one obtains simulated values targeting $\pi$ according to  $\mathbf{q}=\mathbf{m}+\mathbf{S}\bar{\mathbf{q}}$.

Before proceeding, we introduce the following notation. In what follows, $\nabla_{\mathbf{x}}g(\mathbf{x})\in\mathbb{R}^{n}$ denotes the gradient of a scalar valued function  $g(\mathbf{x})$ with
respect to the vector $\mathbf{x}\in\mathbb{R}^{n}$ and $\nabla_{\mathbf{x}} \mathbf{h}(\mathbf{x})\in\mathbb{R}^{m\times n}$  the Jacobian of a vector-valued function $\mathbf{h}(\mathbf{x})\in\mathbb{R}^{m}$. For a function $\mathbf{v}(t)$ of time $t$, the time-derivative $d\mathbf{v}(t)/dt$ is denoted
by $\dot{\mathbf{v}}(t)$. $\mbox{diag}(\cdot)$ is the function that sets to 0 all off-diagonal elements of a matrix and $\mbox{vec}(\cdot)$ the operator that stacks the columns of a matrix into a column vector. $\mathbf{0}_{m}$ denotes the $m$-dimensional zero vector and  $\mathbf{I}_{m}$ the $(m\times m)$ identity matrix. ${\cal N}(\mathbf{x}|\boldsymbol{\mu},\boldsymbol{\Sigma})$ is used to denote the density function of a $N(\boldsymbol{\mu},\boldsymbol{\Sigma})$-distributed random vector $\mathbf{x}$.

\subsection{Generalized randomized HMC (GRHMC) processes\label{subsec:GRHMC}}
GRHMC processes considered by \cite{kleppe_CTHMC} belong to the class of  continuous-time piecewise-deterministic
Markov processes (PDMP) \citep{davis_PDP_book,fearnhead2018} with a dynamics which is defined by an Hamiltonian system.  If a GRHMC process is designed such that its dynamics  results in  a stationary distribution which coincides with the target distribution
and that it can be properly simulated with a sufficiently high precision, it can be used for an (MC)MC  analysis of the target.

The Hamiltonian for a GRHMC process targeting the standardized distribution  $\bar{\pi}(\bar{\mathbf{q}})$ is
\begin{equation}
\mathcal{\mathcal{H}}(\bar{\mathbf{q}},\bar{\mathbf{p}})=-\log\bar{\pi}(\bar{\mathbf{q}})+\frac{1}{2}\bar{\mathbf{p}}^{T}\bar{\mathbf{p}},\label{eq:Hamiltonian_standard}
\end{equation}
where $\bar{\mathbf{q}}$ represents the position of the Hamiltonian system and
$\bar{\mathbf{p}}\in\mathbb{R}^{d}$ is a fictitious momentum variable, which is simulated  at certain random event times $\{t_i\}_i$ (refresh times). Between the refresh times, the   GRHMC process $\bar{\mathbf{z}}(t)=[\bar{\mathbf{q}}(t)^{T}$, $\bar{\mathbf{p}}(t)]^{T},t\in[0,\infty)$
moves deterministically according to the Hamilton's equations associated with the Hamiltonian (\ref{eq:Hamiltonian_standard}), which are given by
\begin{equation}
\dot{\bar{\mathbf{q}}}(t)=\bar{\mathbf{p}}(t),\qquad\dot{\bar{\mathbf{p}}}(t)=\nabla_{\bar{\mathbf{q}}}\log\bar{\pi}\left(\bar{\mathbf{q}}(t)\right),\label{eq:Hamiltons_eq_standard}
\end{equation}
and the momentum  refresh times $\{t_i\}_i$ are determined by  a non-homogeneous Poisson process. 
Here we consider  GRHMC processes where  the  rate of this Poisson process depends on the position, $\lambda = \lambda(\bar{\mathbf{q}})$, and
the momentum is refreshed according to a $N(\mathbf{0}_{d},\mathbf{I}_{d})$-distribution. This defines the baseline GRHMC which we will extend for applications to constrained target distributions (For alternative specifications of GRHMC processes, see \citealp{kleppe_CTHMC}.)
As shown in \cite{kleppe_CTHMC}, such a  baseline GRHMC process
admits $\bar{\pi}(\bar{\mathbf{z}})\propto\exp\{-\mathcal{\mathcal{H}}(\bar{\mathbf{q}},\bar{\mathbf{p}})\}$
as a stationary distribution with  $\bar{\pi}(\bar{\mathbf{q}})$ as the marginal distribution for  $\bar{\mathbf{q}}$.
In the case of a constant Poisson rate
$\lambda$, this GRHMC  reduces to a Randomized HMC, which is geometrically ergodic under weak regularity conditions \citep{bou-rabee2017}.

Suppose a trajectory of the continuous-time process $\bar{\mathbf{q}}(t)$ for the time interval (after the burn-in period) $[0,T]$  has been simulated. Then the resulting values for $\mathbf{q}(t)$ at $N$ discrete, equally spaced times given by
$\mathbf{q}_{(s)}=\mathbf{m}+\mathbf{S}\bar{\mathbf{q}}([s-1]T/[N-1])$, $s=1,\dots,N$, can be used as a regular MCMC sample for approximating the original target  $\pi$. Corresponding MCMC estimates for the expectations $E_{\pi}[\mathcal{M}_{k}(\mathbf{q})]$ of
some scalar monitoring functions $\mathcal{M}_{k}$, $k=1,\dots,K$ with respect to $\pi$ are the discrete-time sample averages $N^{-1}\sum_{s}\mathcal{M}_{k}(\mathbf{q}_{(s)})$. Alternatively,  the expectations $E_{\pi}[\mathcal{M}_{k}(\mathbf{q})]$ can be estimated by the time-integrated averages resulting from the complete simulated trajectory $\bar{\mathbf{q}}(t)$,
\begin{equation}
\hat{r}_{k}=\frac{1}{T}\int_{0}^{T}\mathcal{M}_{k}(\mathbf{m}+\mathbf{S}\bar{\mathbf{q}}(t))dt.\label{eq:time-integrated_averages}
\end{equation}
Time integrated-averages $\hat{r}_{k}$ can have substantially smaller MC standard deviations than averages based on discrete-time samples, especially when large time intervals are required to avoid excessive autocorrelation in the discrete-time samples.

The vector with all variables of the GRHMC process  for estimating expectations of  $K$ monitoring functions is
$\mathbf{y}(t)=[\bar{\mathbf{q}}^{T}(t),\bar{\mathbf{p}}^{T}(t),\Lambda(t),\mathbf{R}^{T}(t)]^{T}$, where $\Lambda(t)$ is
the time-integrated momentum refresh rate defined by $\Lambda(t)=\int_{t_{i-1}}^{t}\lambda(\bar{\mathbf{q}}(\tau))d\tau$, $t\in(t_{i-1},t_{i})$ and $\mathbf{R}^{T}(t)=[\mathbf{R}_{1}(t),\dots,\mathbf{R}_{K}(t)]^{T}$ is the vector of the $K$ time-integrated monitoring functions given by
$\mathbf{R}_{k}(t)= \int_0^t \mathcal{M}_{k}(\bar{\mathbf{q}}(\tau))d\tau$, $k=1,\ldots, K$ with $\hat{r}_{k}=\mathbf{R}_{k}(T)/T$. The behavior of the variables over the entire interval $[0,T]$ is summarized as follows:
\begin{itemize}
\item[$\cdot$]During the interval between two momentum refreshes, $\mathbf{y}(t)$ moves deterministically, where the dynamics of  $\bar{\mathbf{q}}(t)$ and $\bar{\mathbf{p}}(t)$ are given by the ordinary differential equations (ODE) in   Equation (\ref{eq:Hamiltons_eq_standard}) and that of $\Lambda(t)$ and $\mathbf{R}(t)$ by the ODEs
\begin{equation}
\dot{\Lambda}(t) =\lambda(\bar{\mathbf{q}}(t)),\qquad  \dot{\mathbf{R}}_{k}(t)  =\mathcal{M}_{k}(\bar{\mathbf{q}}(t)),\quad k=1,\dots,K.\label{eq:basic_ODE34}
\end{equation}
\item[$\cdot$] The time of the $i$th momentum refresh $t_i$ is determined according to
\begin{equation}
\Lambda(t_{i}-)=u_{i},\quad u_{i}\sim \mbox{Exp}(1),\label{eq:event_eq}
\end{equation}
where $u_i$ is a realization of an exponentially distributed random variable with E$(u_i)=1$.
\item[$\cdot$] At time $t_i$, the states of the variables are updated as follows: $\bar{\mathbf{q}}(t_i)=\bar{\mathbf{q}}(t_{i}-)$,
$\bar{\mathbf{p}}(t_i)\sim N(\mathbf{0}_{d},\mathbf{I}_{d})$, $\Lambda(t_{i})=0$, $\mathbf{R}(t_i)=\mathbf{R}(t_{i}-)$.
\end{itemize}
The values used to initialize the variables' processes are  $\bar{\mathbf{q}}(0)=\bar{\mathbf{\mathbf{q}}}_{0}$,
$\bar{\mathbf{p}}(0)\sim N(\mathbf{0}_{d},\mathbf{I}_{d})$, $\Lambda(0)=0$
and $R_{k}(0)=0$, where $\bar{\mathbf{\mathbf{q}}}_{0}$ is a vector with  some arbitrary value, e.g.~the final state of $\bar{\mathbf{q}}$ in the burn-in period.

\subsection{Numerical generalized randomized HMC (NGRHMC) processes \label{subsec:NGRHMC}}
For all but very simple target distributions $\pi$ (like a standard Gaussian target), the deterministic dynamics (\ref{eq:Hamiltons_eq_standard}) and  (\ref{eq:basic_ODE34}) in the interval  between two momentum refreshes does not admit closed-form solutions so that numerical integrators for approximative solutions are required. Here we use an adaptive step size Runge-Kutta procedure with a pair of integrators of order 3(2) as proposed by \citet{BOGACKI1989321}
(BS) \citep[see also][]{doi:10.1137/S1064827594276424}. The approximations of such an adaptive step-size
integrator are typically biased relative to the exact dynamics of the GRHMC.
However, this bias can be  controlled by the integrator's error control mechanism  and, as demonstrated in \cite{kleppe_CTHMC} (and also in Section \ref{subsec:Illustrative-examples} here), produces biases in subsequent (MC)MC estimates which are typically negligible relative to their total MC variation  even with a generous integration error tolerance.

By tolerating a small bias and using an adaptive step-size integration, one obtains a numerically very robust algorithm and circumvents the difficult tuning of the necessarily fixed
step sizes used by  standard symplectic integrators.

The adaptive step-size Runge-Kutta BS integrator approximates  the ODE state vector $\mathbf{y}(t)$ at discrete mesh time points $\{\tau_{j}\}_{j}$,
where the  time step-sizes $\tau_{j+1}-\tau_{j}>0$  with $\tau_{0}=0$  are determined by the error control mechanism \citep{10.5555/153158,numrecipes2007}.
The resulting approximations to  $\mathbf{y}(\tau_{j})$ are denoted by  $\hat{\mathbf{y}}_{j}$. To approximate $\mathbf{y}(t)$ at non-mesh points, a cubic Hermite interpolation 
formula (which is third-order accurate) \citep{10.5555/153158} is used, and the resulting numerically
integrated ODE state vector (and hence the value of the numerical GRHMC process) is denoted by $\hat{\mathbf{y}}_{j}(t)$ for $t\in[\tau_{j},\tau_{j+1}]$.

To detect and locate the event of a momentum update within a Runge-Kutta BS time step from $\tau_j$ to $\tau_{j+1}$, we resort to a root-finding algorithm embedded in the Runge-Kutta method. Such an algorithm searches for a $t\in (\tau_j,\tau_{j+1})$  that solves
\begin{equation}
\text{\ensuremath{\gamma}}(\hat{\mathbf{y}}_{j}(t))=0,\label{eq:event_roots}
\end{equation}
where $\gamma$ represents an  indicator function that indicates the occurrence of a certain event in the process $\hat{\mathbf{y}}_{j}(t)$.
The specific form of $\gamma$  for the event of a momentum refresh is according to Equation  (\ref{eq:event_eq}) given by $\gamma(\mathbf{y}(t))=\Lambda(t)-u_i$. If the root finding algorithm indicates that there is a momentum refresh in the NGRHMC process, say at time $t_{m}\in(\tau_{j},\tau_{j+1})$, then the Runge-Kutta BS step is truncated by moving the right mesh point to $t_m$, i.e.~$\tau_{j+1}\leftarrow t_{m}$. This is then followed by updating the numerically integrated states $\hat{\mathbf{y}}$ at time $\tau_{j+1}=t_{m}$ according to the momentum refresh mechanism described in Section \ref{subsec:GRHMC} above.
In our NGRHMC processes for constrained target distributions presented below, not only the event of a momentum refresh needs to be monitored,  but also the event that $\mathbf{q}$ reaches the boundary of its restricted domain, which then also require a truncation of the integrator step.
Therefore, in these processes, root finding algorithms based on  Equation (\ref{eq:event_roots}) will also be instrumental for detecting and locating the latter type of events.

Before we proceed, it is worth noting that the  Runge-Kutta BS integrator of order 3(2) that we use for the baseline NGRHMC and also for its extension for constrained targets  differs from the integrator suggested by  \citet{kleppe_CTHMC}.
He proposes to use an adaptive step size Runge-Kutta-Nystr{\o}m method \citep[see e.g.~][]{DORMAND1987937,10.5555/153158} of order 6(5) for a second-order ODE representation of $\dot{\bar{\mathbf{q}}}(t)$ and $\dot{\bar{\mathbf{p}}}(t)$ in Equation (\ref{eq:Hamiltons_eq_standard}) combined with first-order rules for $\dot{\Lambda}(t)$ and $ \dot{\mathbf{R}}_{k}(t)$ in Equation (\ref{eq:basic_ODE34}). One reason for using the lower-order Runge-Kutta BS integrator and not the higher-order integrator recommended by \citet{kleppe_CTHMC} is that it may happen that the NGRHMC process for constrained targets  $\mathbf{q}$  quite often collides with the boundary so that many of the integrator steps need to be truncated.
Since the  time steps of lower-order integrators are typically shorter and require  less computing time  per step compared to higher-order integrators, the use of a low-order integrator avoids  significant parts of large steps being discarded when the step needs to be truncated and thus  saves overall computation time. 
A second reason for using a low-order integrator is that we use for the interpolation between two Runge-Kutta BS integrator steps a third order Hermite polynomial which, as discussed further below, simplifies solving the event root equation (\ref{eq:event_roots}) for certain  conditions on $\gamma$.
In principle, the higher order Runge-Kutta-Nystr{\o}m integrator could also be combined with this third-order polynomial for interpolation. However,  the use of this low order interpolation scheme would then counteract the effects of the increased accuracy  that can be achieved when moving  from a computationally cheap low order integrator to a computationally costly higher order one.
A third reason for using a  first-order ODE solver such as the Runge-Kutta BS integrator is that such a scheme can be directly  carried  over to a Riemann manifold version \citep{2211.01746} of our proposed constrained-target NRGHMC  which we intend to explore in future research.

Finally, it should be mentioned that many available codes for  numerical integration of ODE processes have subroutines for root finding algorithms, so that they could be used for an implementation of NGRHMC processes including our proposed processes for constrained targets  without much programming effort. An example for such a code is the R function \texttt{lsodar} written by \cite{Hindmarsh1983}, which is available in the R-package deSolve of \citealp{Soetaert2010-JSS-deSolve}, and was used for NGRHMC purposes by \cite{2403.07495}.
However, to achieve the best possible numerical performance, the adaptive step size Runge-Kutta BS method is implemented in the pdmphmc package along with functions for root finding algorithms that take into account the specific forms of the constraints (as described below).

\section{GRHMC for constrained target distributions\label{sec:contrained_NGRHMC}}
\subsection{Basic principle\label{subsec:Basic-principle}}
Now consider target distributions with constrained support, with density kernel $\tilde{\pi}_c$ of the form
\[
\tilde{\pi}_c(\mathbf{q})=\tilde{\pi}(\mathbf{q})\prod_{r=1}^{R}\mathds{1} [c_{r}(\mathbf{q})],\quad \mathds{1} [c_{r}(\mathbf{q})]=\begin{cases}
1 & \text{if }c_{r}(\mathbf{q})\geq0\\
0 & \text{otherwise},
\end{cases}
\]
where $\tilde{\pi}$
is the density kernel of a non-truncated distribution and  $c_{r}(\mathbf{q})\geq0,\;r=1,\dots,R$ is a set of linear and/or non-linear constraints on $\mathbf{q}$.
The restricted domain defined by $\Omega=\{\mathbf{q}\in\mathbb{R}^{d}:c_{1}(\mathbf{q})\ge0,c_{2}(\mathbf{q})\geq0,\dots,c_{R}(\mathbf{q})\geq0\}$
is assumed to be non-empty and non-degenerate in the sense that
$\Omega$ is not a lower-dimensional manifold such as a plane or a
sphere embedded in $\mathbb{R}^{d}$.
In terms of standardized coordinates $\bar{\mathbf{q}}$, the constraints and the restricted domain are $\bar{c}_{r}(\bar{\mathbf{q}})=c_r(\mathbf{m}+\mathbf{S}\bar{\mathbf{q}})\geq0$  and $\bar{\Omega}=\{\bar{\mathbf{q}}:\mathbf{m}+\mathbf{S}\bar{\mathbf{q}}\in\Omega\}$, respectively, and the density kernel of the standardized truncated target has the form
\begin{equation}
\bar{\tilde{\pi}}_c(\bar{\mathbf{q}})= \bar{\tilde{\pi}}(\bar{\mathbf{q}})\prod_{r=1}^{R}\mathds{1} [\bar{c}_{r}(\bar{\mathbf{q}})],\quad  \bar{\tilde{\pi}}(\bar{\mathbf{q}})\propto\tilde{\pi}(\mathbf{m}+\mathbf{S}\bar{\mathbf{q}}),\label{eq:truncted_stand_target}
\end{equation}
where $\bar{\tilde{\pi}}(\bar{\mathbf{q}})$ represents the kernel of the  non-truncated standardized density.

An easy to implement approach to sampling from such a constrained distribution  is to draw $\bar{\mathbf{q}}$'s from the unconstrained distribution $\bar{\tilde{\pi}}$
and then discard the draws  that violate  the constraints.
However, in situations where the probability of rejecting a draw (i.e.~the probability under $\bar{\tilde{\pi}}$ that $\bar{\mathbf{q}}\notin \bar{\Omega}$), is close to one, the computing time for a (MC)MC analysis of the constrained target distribution can become prohibitively long.
This  can easily happen, for example, in applications with high-dimensional distributions in which the domain of many or all elements in $\bar{\mathbf{q}}$ is restricted.
In contrast to this brute force approach, our NGRHMC method for constrained targets   is designed  to explore with the simulated $\bar{\mathbf{q}}$ trajectories only the relevant domain $\bar\Omega$.
Furthermore, as illustrated in our applications presented below, the brute force approach is also not  suited to situations where parameter constraints are used to identify the parameters of a Bayesian model in which the unconstrained posterior distribution of the parameters has a complicated multimodal form.

To construct a GRHMC process for the constrained standardized target (\ref{eq:truncted_stand_target}),
one can use the regular GRHMC of Section \ref{subsec:GRHMC} for the behavior of $\bar{\mathbf{q}}$ in the interior of $\bar{\Omega}$ and then combine this with a mechanism that ensures that $\bar{\mathbf{q}}$ remains in $\bar{\Omega}$. This can be achieved by an appropriate update of the momentum $\bar{\mathbf{p}}$ whenever $\bar{\mathbf{q}}$ collides with the boundary of $\bar{\Omega}$ such that $\bar{\mathbf{q}}$ moves back into $\bar{\Omega}$.
For  targets subject to $R$ constraints, this means  that when $\bar{\mathbf{q}}$ collides with the boundary of the $r$-th constraint, indicated by $\bar{c}_{r}(\bar{\mathbf{q}}(t))=0$, the momentum update must  be generated  by a specific transition kernel, denoted by $K_{r,\bar{\mathbf{q}}}(\bar{\mathbf{p}}|\bar{\mathbf{p}}^{\prime})$, that has the properties that (i) $\bar{\mathbf{p}}(t)\sim K_{r,\bar{\mathbf{q}}}(\cdot|\bar{\mathbf{p}}(t-))$ in Equation (\ref{eq:Hamiltons_eq_standard}) leads to a position  $\bar{\mathbf{q}}(t) \in \bar{\Omega}$ and (ii)
that it leaves the constrained target distribution $\bar{\tilde{\pi}}_c$ invariant.

\citet{BIERKENS2018148} provide sufficient conditions  for  transition  kernels at the boundary of the constrained domain of PDMP processes to have these two properties (see their Equations 4 and 5), which also apply to the GRHMC processes considered here.
These sufficient conditions for the boundary transition kernel $K_{r,\bar{\mathbf{q}}}$ in a GRHMC process are
\begin{equation}
\int K_{r,\bar{\mathbf{q}}}(\bar{\mathbf{p}}|\bar{\mathbf{p}}^{\prime})\mathcal{N}(\bar{\mathbf{p}}^{\prime}|\mathbf{0}_{d},\mathbf{I}_{d})d\bar{\mathbf{p}}^{\prime}=\mathcal{N}(\bar{\mathbf{p}}|\mathbf{0}_{d},\mathbf{I}_{d}),\quad \forall\;\bar{\mathbf{q}}\in\bar{\Omega}:\bar{c}_{r}(\bar{\mathbf{q}})=0,\label{eq:general_preservation_std_normal}
\end{equation}
and
\begin{equation}
\int\left[\bar{\mathbf{p}}^{T}\bar{\mathbf{n}}_{r}(\bar{\mathbf{q}})\right]K_{r,\bar{\mathbf{q}}}(\bar{\mathbf{p}}|\bar{\mathbf{p}}^{\prime})d\bar{\mathbf{p}}=-\left[\bar{\mathbf{p}}^{\prime}\right]^{T}\bar{\mathbf{n}}_{r}(\bar{\mathbf{q}}),
\quad\forall\;\bar{\mathbf{q}}\in\bar\Omega:\bar{c}_{r}(\bar{\mathbf{q}})=0,\text{ and }\bar{\mathbf{p}}^{\prime}\in\mathbb{R}^{d},\label{eq:IP_reversal}
\end{equation}
where $\bar{\mathbf{n}}_{r}(\bar{\mathbf{q}})$ is the inward-pointing normal vector for the $r$-th constraint given by
\begin{equation}
\bar{\mathbf{n}}_{r}(\bar{\mathbf{q}})=\nabla_{\bar{{\mathbf{q}}}}\bar{c}_{r}(\bar{\mathbf{q}}) =\mathbf{S}\left[\nabla_{\mathbf{q}}c_{r}(\mathbf{q})\right]\bigg{|}_{\mathbf{q}=\mathbf{m}+\mathbf{S}\bar{\mathbf{q}}}.\label{eq:normal_vector}
\end{equation}
The condition in Equation (\ref{eq:general_preservation_std_normal}) says that $K_{r,\bar{\mathbf{q}}}$ has the standard normal distribution of the momentum as its invariant distribution and the condition in Equation (\ref{eq:IP_reversal}) means that
the expected inner product of the momentum reversal vector and  the inward-pointing normal vector must be equal to the reflected inner product for the actual momentum vector when hitting the boundary.

\subsection{Boundary transition kernels\label{subsec:Boundary-kernels}}
Existing  linear dynamics PDMP approaches for constrained domains are typically based on boundary transition kernels that reflect the  PDMP process at the boundary \citep[see e.g.~][]{BIERKENS2018148}.
When such  boundary reflections are used for Hamiltonian dynamics, this results in a collision behavior called 'inelastic collision'.
This type of collisions in a discrete-time HMC approach for the special case of truncated Gaussian target distributions can be found in \citet{doi:10.1080/10618600.2013.788448}.

For the GRHMC processes considered here, such  collisions would arise for the following deterministic boundary transition kernel:
\begin{equation}
K_{r,\bar{\mathbf{q}}}(\bar{\mathbf{p}}|\bar{\mathbf{p}}^{\prime})=\delta_{ [\mathbf{F}_{\bar{\mathbf{q}}}\bar{\mathbf{p}}^{\prime}]}(d\bar{\mathbf{p}}), \quad\text{where }\;\mathbf{F}_{\bar{\mathbf{q}}}=\mathbf{I}_{d}-2\mathbf{R}_{\bar{\mathbf{q}}},\quad\mathbf{R}_{\bar{\mathbf{q}}}=\left([\bar{\mathbf{n}}_{r}(\bar{\mathbf{q}})]^{T}[\bar{\mathbf{n}}_{r}(\bar{\mathbf{q}})]\right)^{-1}[\bar{\mathbf{n}}_{r}(\bar{\mathbf{q}})][\bar{\mathbf{n}}_{r}(\bar{\mathbf{q}})]^{T}
\label{eq:collition_transition_kernel},
\end{equation}
which produces  momentum updates that can be written as
\begin{equation} \bar{\mathbf{p}}=\mathbf{F}_{\bar{\mathbf{q}}}\bar{\mathbf{p}}^{\prime}=\bar{\mathbf{p}}^{\prime}-2\left(\frac{\bar{[\mathbf{p}}^{\prime}]^{T}\bar{\mathbf{n}}_{r}(\bar{\mathbf{q}})}{[\bar{\mathbf{n}}_{r}(\bar{\mathbf{q}})]^{T}\bar{\mathbf{n}}_{r}(\bar{\mathbf{q}})}\right)\bar{\mathbf{n}}_{r}(\bar{\mathbf{q}}).
\label{eq:collition_transition_kernel_alt}
\end{equation}
This standard transition kernel leaves the standard normal distribution of the momentum invariant since $\mathbf{F}_{\bar{\mathbf{q}}}\mathbf{F}_{\bar{\mathbf{q}}}^{T}=\mathbf{I}_{d}$, and thus satisfies
condition (\ref{eq:general_preservation_std_normal}).
Furthermore, $\bar{\mathbf{p}}^{T}[\bar{\mathbf{n}}_{r}(\bar{\mathbf{q}})]=\left[\bar{\mathbf{p}}^{\prime}\right]^{T}\mathbf{F}_{\bar{\mathbf{q}}}^{T}[\bar{\mathbf{n}}_{r}(\bar{\mathbf{q}})]=-\left[\bar{\mathbf{p}}^{\prime}\right]^{T}[\bar{\mathbf{n}}_{r}(\bar{\mathbf{q}})]$, so that $\bar{\mathbf{q}}$ is prevented from leaving $\bar \Omega$  according to condition (\ref{eq:IP_reversal}).

\begin{figure}
\begin{centering}
\includegraphics[scale=0.5]{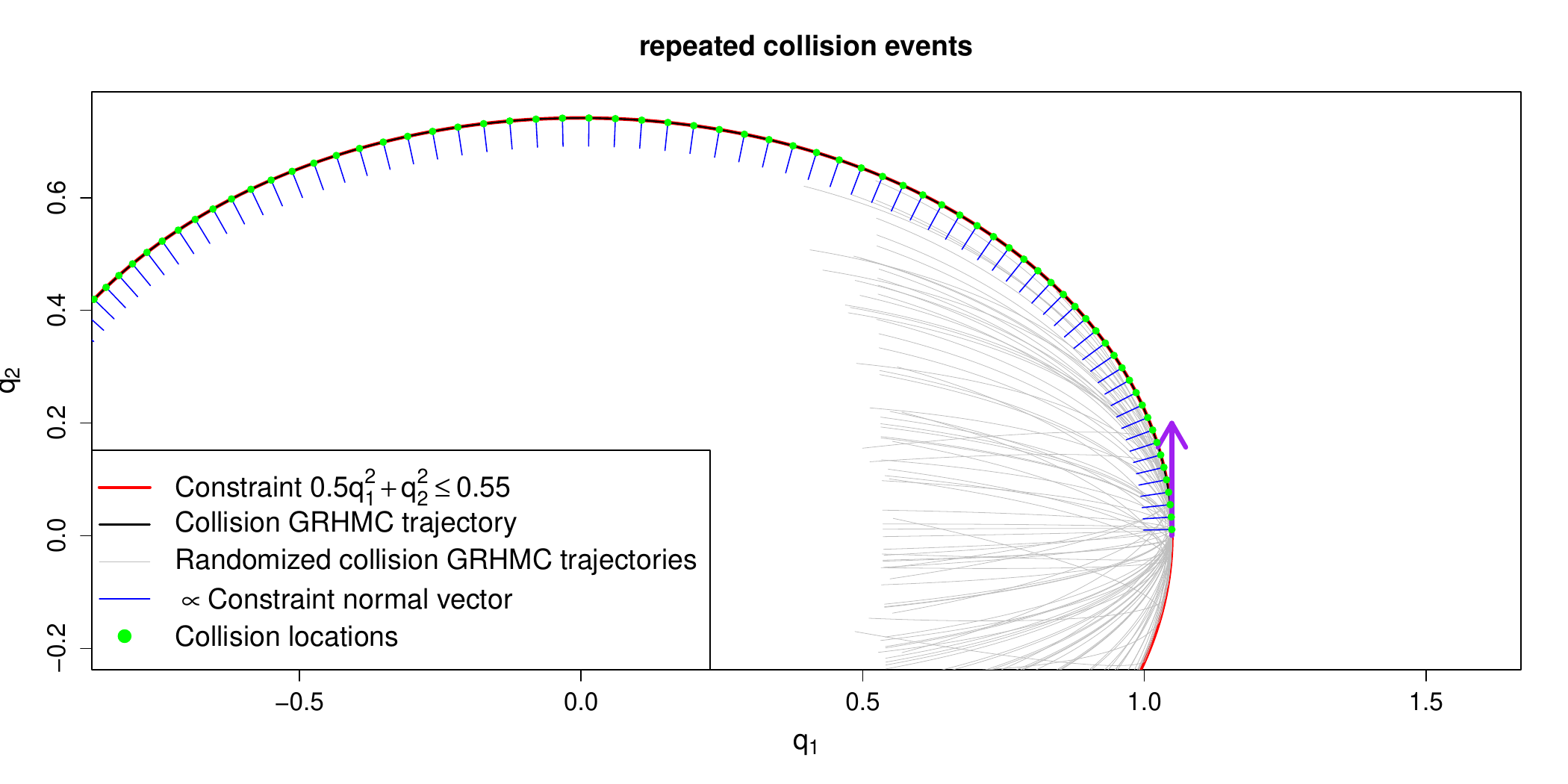}
\par\end{centering}
\centering{}\caption{\label{fig:Repeated-collision-events}Repeated collision events for
the  boundary transition kernel (\ref{eq:collition_transition_kernel}) applied to a bivariate standard
Gaussian target subject to the constraint $0.55- 0.5q_{1}^{2}- q_{2}^{2}\protect\geq0$.
The purple arrow indicates the initial location and direction of travel of the process trajectory.
The simulation was stopped after 100 collisions.  The thin gray lines are the initial part of the trajectories obtained
when using a randomized boundary transition kernel.}
\end{figure}

A disadvantage of the deterministic  transition kernel (\ref{eq:collition_transition_kernel}) is that in cases where $\bar \Omega$ is strongly convex (at least locally), repeated boundary collisions can occur with a high and increasing frequency.
This behavior is due to the fact that for this boundary transition
kernel, the angle between $\bar{\mathbf{n}}_r$ and $\bar{\mathbf{p}}$
is the same as between $\bar{\mathbf{n}}_r$ and $-\bar{\mathbf{p}}^{\prime}$,
since $\parallel\bar{\mathbf{p}}\parallel=\parallel\bar{\mathbf{p}}^{\prime}\parallel$.
Such a situation is illustrated in Figure \ref{fig:Repeated-collision-events}
for a bivariate standard Gaussian distribution  $\tilde{\pi}$ subject to the constraint
$0.55-0.5q_{1}^{2}-q_{2}^{2}\geq 0$. The consequence of this boundary collision behavior is that numerical integrators for the resulting constrained GRHMC process are computationally costly since each period between two collisions must contain at least one integrator time step.
As an alternative to the deterministic transitions kernel (\ref{eq:collition_transition_kernel}), we therefore propose to use a randomized version of this transition kernel that is more resistant to such an undesirable collision behavior.

\subsection{Randomized boundary transition kernels}
The proposed randomized  transition kernel for the momentum update of the GRHMC process at the boundary, denoted by $\tilde{K}_{r,\bar{\mathbf{q}}}$,  modifies  the standard kernel (\ref{eq:collition_transition_kernel}) by randomizing its deterministic  momentum updates (\ref{eq:collition_transition_kernel_alt}) according to
\begin{equation}
\bar{\mathbf{p}}=\mathbf{z}-\frac{(\bar{\mathbf{p}}^{\prime}+\mathbf{z})^{T}[\bar{\mathbf{n}}_{r}(\bar{\mathbf{q}})]}{[\bar{\mathbf{n}}_{r}(\bar{\mathbf{q}})]^{T}[\bar{\mathbf{n}}_{r}(\bar{\mathbf{q}})]}[\bar{\mathbf{n}}_{r}(\bar{\mathbf{q}})],\quad \mathbf{z}\sim N(\mathbf{0}_{d},\mathbf{I}_{d}).\label{eq:randomized_boundary_kernel}
\end{equation}
Such a randomized boundary transition based on $(d-1)$-dimensional standard normal distributions on the space orthogonal to $\bar{\mathbf{n}}_{r}(\bar{\mathbf{q}})$  is mentioned in  \citet{2303.08023} (see also \citealp{1706.04781}) as a possible option for constrained PDMP processes, but is not examined in detail there.

As shown in  the Appendix,
the randomized transition kernel $\tilde{K}_{r,\bar{\mathbf{q}}}$ as defined in Equation (\ref{eq:randomized_boundary_kernel}) satisfies the conditions (\ref{eq:general_preservation_std_normal}) and  (\ref{eq:IP_reversal}), ensuring its validity.
Furthermore, it generates momentum updates $\bar{\mathbf{p}}$  that do not preserve the momentum norm of $\bar{\mathbf{p}}^{\prime}$, resulting in an angle between
$\bar{\mathbf{n}}_r$ and $\bar{\mathbf{p}}$ that differs from that for $\bar{\mathbf{n}}_r$ and $-\bar{\mathbf{p}}^{\prime}$.
This prevents the rapid succession of collisions along the boundary produced by the deterministic transition kernel (\ref{eq:collition_transition_kernel}) illustrated in Figure \ref{fig:Repeated-collision-events}.
In this figure, we also provide the first part of 100 trajectories for the position variable generated by the randomized transition kernel based on the same initial position and momentum used for the deterministic kernel.
It can be seen that the trajectories generated by the randomized transition kernel systematically avoid the repeated collisions.

Finally, it should be noted that the specification of the proposed randomized transition kernel (\ref{eq:randomized_boundary_kernel}) is by no means the only possible way to specify  a valid randomized kernel.
Possible alternatives could be, for example, a combination of the randomized transition kernel (\ref{eq:randomized_boundary_kernel})
with either the deterministic kernel (\ref{eq:collition_transition_kernel_alt}) or $-\bar{\mathbf{p}}^{\prime}$ as respective components of a mixture of distributions. However, our proposed randomized transition kernel has the advantages that it is  easy to implement, computationally fast, does not require setting any tuning parameters, and works well in the implementation considered below.  We therefore refrain from pursuing such alternatives further here.

\subsection{Sparsity\label{subsec:sparsity}}
In applications where the target distribution is constrained, often only a small subset of the elements in $\bar{\mathbf{q}}$ are affected by the restriction, so that the inward-pointing normal vector $\bar{\mathbf{n}}_{r}(\bar{\mathbf{q}})$ is a sparse vector with many zeros.
However, by design, the transition kernel (\ref{eq:randomized_boundary_kernel}) updates all elements in $\bar{\mathbf{p}}$, where the elements of $\bar{\mathbf{p}}$ that correspond to the zero elements in $\bar{\mathbf{n}}_{r}(\bar{\mathbf{q}})$ are updated by a draw from an N(0,1) distribution.
If boundary collisions occur relatively frequently, this can lead to a suboptimal random walk-like behavior of the variables in $\bar{\mathbf{q}}$ that are not affected by the constraint.
Note that the deterministic kernel (\ref{eq:collition_transition_kernel}) also has this sparsity property, since $[\mathbf{F}_{\bar{\mathbf{q}}}\bar{\mathbf{p}}^{\prime}-\bar{\mathbf{p}}^{\prime}]_{i}=0$
whenever $\left[\bar{\mathbf{n}}_{r}(\bar{\mathbf{q}})\right]_{i}=0$.

A simple solution to largely avoid such suboptimal behavior is to modify the randomized transition kernel defined by Equation (\ref{eq:randomized_boundary_kernel}) so that  it only updates  the elements in $\bar{\mathbf{p}}$ that correspond to the non-zero elements in $\bar{\mathbf{n}}_{r}(\bar{\mathbf{q}})$.
To represent this modification, let $\mathcal{A}(r)$ denote the  index set, which contains the indices of those elements in $\bar{\mathbf{q}}$  that are affected by the $r$th restriction $\bar{c}_r(\bar{\mathbf{q}}) \geq 0$, i.e.~$\mathcal{A}(r)=\{i: \partial \bar{c}_{r}(\bar{\mathbf{q}})/\partial \bar{q}_{i}\neq 0, \bar{\mathbf{q}}\in\bar{\Omega},i=1,\ldots,d\}$,
with the corresponding complementary index set denoted by $-\mathcal{A}(r)$. Furthermore, let $|\mathcal{A}(r)|$ be the number of elements in $\mathcal{A}(r)$.
Then the momentum updates according to the `sparse' version   of the randomized transition kernel, denoted by $\tilde{K}^{S}_{r,\bar{\mathbf{q}}}$,  are given by
\begin{align}
&\qquad\qquad\qquad\qquad\qquad\qquad\bar{\mathbf{p}}=( \bar{\mathbf{p}}_{\mathcal{A}(r)},\bar{\mathbf{p}}_{-\mathcal{A}(r)}),\quad \mbox{with} \label{eq:sparse_randomized_boundary_kernel}\\
&\bar{\mathbf{p}}_{-\mathcal{A}(r)}=\bar{\mathbf{p}}_{-\mathcal{A}(r)}^{\prime},\quad
\bar{\mathbf{p}}_{\mathcal{A}(r)}= \mathbf{z}^S-\frac{(\bar{\mathbf{p}}_{\mathcal{A}(r)}^{\prime}+\mathbf{z}^S)^{T}[\bar{\mathbf{n}}_{r}(\bar{\mathbf{q}})]_{\mathcal{A}(r)}}{[\bar{\mathbf{n}}_{r}(\bar{\mathbf{q}})]^{T}[\bar{\mathbf{n}}_{r}(\bar{\mathbf{q}})]}[\bar{\mathbf{n}}_{r}(\bar{\mathbf{q}})]_{\mathcal{A}(r)},\quad
\mathbf{z}^S\sim N(\mathbf{0}_{|\mathcal{A}(r)|},\mathbf{I}_{|\mathcal{A}(r)|}), \nonumber
\end{align}
where the vectors indexed by $\mathcal{A}(r)$ and $-\mathcal{A}(r)$ are the corresponding subvectors obtained by eliminating all elements  with an index not contained in $\mathcal{A}(r)$ and $-\mathcal{A}(r)$, respectively.
As can be easily shown, this adaption of the randomized transition kernel (\ref{eq:randomized_boundary_kernel}) leaves its validity unaffected.
When implementing the adapted sparse transition kernel (\ref{eq:sparse_randomized_boundary_kernel}) for cases where only a few elements are affected by the constraint, one can also  take advantage of the fact that the constraint representations involve sparse matrices, so that sparse matrix functions and sparse storage schemes can be used  for reducing computing time.

\subsection{Pseudo-Code of the constrained  NGRHMC process}
\begin{algorithm}
\begin{algorithmic}
\State Select values for the  tuning parameters $\mathbf{m}$, $\mathbf{S}$, $T$ and $N$.
\vspace{0.1cm}
\State $t \gets 0$, $\bar{ \mathbf q}(0)\gets \bar{\mathbf q}_0$, $\bar{\mathbf p}(0)\sim N(\mathbf 0_d, \mathbf{I}_d)$, $u \sim \text{Exp}(1)$, $s\gets 1$.
\While{$t < T$}
\State Propose integrator step size $\varepsilon$.
\State Perform a Runge-Kutta BS step with step size $\varepsilon$ for the ODEs  (\ref{eq:Hamiltons_eq_standard}) and (\ref{eq:basic_ODE34}) from $\mathbf y(t)$ to $\mathbf y(t+\varepsilon)$.
\State Determine if some $h_\lambda \in (0,1)$ so that $\Lambda(t+h_\lambda\varepsilon)=u$ exist. Otherwise set $h_\lambda=1$.
\State For $r=1,\dots,R$, determine if $h_r \in (0,1)$ so that  $\bar c_r(\bar{ \mathbf q}(t+h_r \varepsilon))=0$ exist. Otherwise set $h_r=1$.
\State $h \gets \min(h_\lambda,h_1,\dots,h_R)$
\While{$(s-1)T/(N-1)<t+h\varepsilon$} \hspace{0.5cm}  /\!/ {\sl Store any discrete time samples in time interval $(t,t+h\varepsilon)$}
\State $\mathbf q^{(s)} \gets \mathbf m+\mathbf S \bar{\mathbf q}((s-1)T/(N-1))$ and $s\gets s+1$.
\EndWhile
\If{h==1}  \hspace{0.5cm} /\!/ {\sl No event occurred that requires a momentum update}
\State $t \gets t+\varepsilon$
\Else      \hspace{0.5cm}/\!/   {\sl Event occurred that requires a momentum update}
\State $\bar{\mathbf p}^* \gets \bar{\mathbf p}(t+h\varepsilon)$
\State $t \gets t+h\varepsilon$
\If{$h_\lambda<\min(h_1,\dots,h_R)$} \hspace{0.5cm}  /\!/ {\sl Regular momentum refreshment}
\State $\bar{\mathbf p}(t)\sim N(\mathbf 0_d, \mathbf{I}_d)$.
\State $u \sim \text{Exp}(1)$.
\State $\Lambda(t) \gets 0$.
\Else  \hspace{0.5cm}  /\!/ {\sl Boundary collision momentum update}
\State $\hat r = \arg \min_{r\in (1,\dots,R)} h_r$
\State $\bar{\mathbf p}(t) \sim \tilde{K}_{\hat r,\bar{\mathbf q}(t+h\varepsilon)}(\cdot|\bar{\mathbf p}^*)$.
\EndIf
\EndIf
\EndWhile
\State \textbf{Return} $\mathbf q^{(s)},\;s=1,\dots,N$ and $\mathbf R(T)/T$.
\end{algorithmic}
\caption{\label{alg:Restricted-domain-NGRHMC} Constrained domain NGRHMC algorithm.
}
\end{algorithm}
To numerically integrate and interpolate the GRHMC process with the proposed  (sparse) randomized boundary transition kernel for constrained target distributions, as generically presented in Sections (\ref{subsec:Basic-principle})-(\ref{subsec:sparsity}),  we rely on the Runge-Kutta BS integrator and the Hermite interpolation scheme, which are described in Section (\ref{subsec:NGRHMC}). The event indicator function $\gamma$ in Equation (\ref{eq:event_roots}) for locating a collision of the NGRHMC process with the boundary of the $r$th constraint has the form $\gamma_r(\mathbf{y}(t))=\bar{c}_{r}(\bar{\mathbf{q}}(t))$, $r=1,\ldots,R$. Thus, together with the event indicator function for a regular momentum update, $\gamma(\mathbf{y}(t))=\Lambda(t)-u_i$, the constrained NGRHMC algorithm requires monitoring a total of $R+1$ indicator functions.
Note that if multiple event indicators have roots that solve Equation (\ref{eq:event_roots}) within an integrator time step,  the constrained NGRHMC algorithm needs to select the earliest of these to then update the momentum accordingly, either as a regular refreshment or as a boundary-collision update.

To summarize the presentation of the proposed constrained domain NGRHMC approach (pending discussion of further implementation details for various specific types of constraints in the next section), we provide a pseudo-code of it in Algorithm 1. For simplicity, the pseudo-code does not distinguish between the exact ODE flow $\mathbf{y}(t)$ and the numerically integrated
and interpolated counterparts $\hat{\mathbf{y}}_j$ and  $\hat{\mathbf{y}}_j(t)$.

\section{Types of constraints}
Here we consider different types of the constraints $c_{r}(\mathbf{q}) \geq 0$ and outline how their specific structure can be exploited to simplify the search of event roots associated with Equation (\ref{eq:event_roots}) and the calculation  of the normal vector $\mathbf{n}_{r}(\mathbf{q})$ in Equation (\ref{eq:normal_vector}) for the constrained NGRHMC. All these types of constraints are implemented in the pdmphmc package.
\subsection{Linear constraints\label{subsec:Linear-constraints}}
In the case where the non-standardized target distribution $\tilde \pi(\mathbf{q})$ is subject to linear constraints of the form
\[
c_{r}(\mathbf{q})=\mathbf{a}_{r}^{T}\mathbf{q}+b_{r}\geq0,
\]
where  $\mathbf{a}_{r}$ and $b_{r}$ are fixed coefficients, the corresponding constraints for the standardized target $\bar{\pi}(\bar{\mathbf{q}})$ obtain as
$\bar{c}_{r}(\bar{\mathbf{q}})=\bar{\mathbf{a}}_{r}^{T}\bar{\mathbf{q}}+\bar{b}_{r}\geq0$, where $\bar{\mathbf{a}}_{r}=\mathbf{S}\mathbf{a}_{r}$ and $\bar{b}_{r}=\mathbf{a}_{r}^{T}\mathbf{m}+b_{r}$, and
the inward-pointing normal vector simplifies to $\bar{\mathbf{n}}(\bar{\mathbf{q}})=\bar{\mathbf{a}}_{r}$.
For such a linear constraint, the cubic Hermite polynomial for the interpolation between two integrator steps leads to  root equations for locating  boundary collisions,  which consist of  3rd order polynomial equations with closed form solutions.
When using the pdmphmc package, the user only needs to provide a procedure for evaluating the function $c_{r}$.
pdmphmc supports both sparse and dense storage for $\mathbf{a}_{r}$.

\subsection{Norm constraints\label{subsec:norm-constraints}}
The next type of constraints for  $\tilde \pi(\mathbf{q})$  we consider are ${\ell}_1$-  and ${\ell}_2$-norm constraints on linear functions of $\mathbf{q}$ of the form $\mathbf{w}=\mathbf{A}\mathbf{q}+\mathbf{b}$.

The ${\ell}_1$-norm constraint,   $\parallel\mathbf{w}\parallel_{1}\leq\text{\ensuremath{v}}$, for some fixed constant $v$ implies that
\[
c_{r}(\mathbf{q})=v-\parallel\mathbf{A}\mathbf{q}+\mathbf{b}\parallel_{1}\geq0.
\]
For the 
implementation of the NGRHMC, we take advantage of the fact that for time intervals in which the sign of none of the elements in $\mathbf{w}(t)=\mathbf{A}\mathbf{q}(t)+\mathbf{b}=\mathbf{A}(\mathbf{m} +\mathbf{S}\bar{\mathbf{q}}(t))+\mathbf{b}$ changes,  $\parallel\mathbf{w}(t)\parallel_{1}$ can be represented as $\mathbf{w}(t)^{T}\text{sign}(\mathbf{w}(t))$. Based on this, the algorithm for locating  boundary collisions is split into two simple root-finding problems. First it searches for times at which the sign of any element of $\mathbf{w}(t)$ changes, and then searches for  boundary collisions within the resulting time intervals between two sign changes.
The root equations for both problems consist  of 3rd order polynomial equations with closed form solutions (analogous to those for locating  collisions with the boundary  of linear restrictions in Section \ref{subsec:Linear-constraints}).
If a collision with the boundary of the ${\ell}_1$-norm constraint is localized at time $t$, then the inward-pointing normal vector $\bar{\mathbf{n}}(\bar{\mathbf{q}})$  is computed  according to $\nabla_{\bar{\mathbf{q}}}\bar{c}(\bar{\mathbf{q}}(t))=-\mathbf{S}\mathbf{A}^{T}\text{sign}(\mathbf{w}(t))$.
The pdmphmc package only requires the user to provide a code that evaluates $\mathbf{w}$, and it uses sparse (compressed row) storage  for $\mathbf{A}$.

The ${\ell}_2$-norm constraint,   $\parallel\mathbf{w}\parallel_{2}\leq\text{\ensuremath{v}}$, implies that
\[
c_{r}(\mathbf{q})=v^{2}-\parallel\mathbf{A}\mathbf{q}+\mathbf{b}\parallel_{2}^{2}\geq0.
\]
Here the cubic Hermite polynomial interpolation between integrator steps  implies that the boundary collision times  are the  roots of polynomial equations of order 6. The existence  of roots indicating boundary collisions  within an integrator step  is checked using Sturm's theorem.
If a collision occurs, the time of its occurrence is determined using Newton's method. The  pdmphmc interface  for the user is otherwise similar to that for the ${\ell}_1$-norm constraint.

\subsection{Constraints of the form $F(\mathbf{A}\mathbf{q}+\mathbf{b})\protect\geq0$\label{subsec:Non-lin-Constraints-on-the} }
The most general type of constraints   we consider  restricts the domain of $\mathbf{q}$ via $\mathbf{w}=\mathbf{A}\mathbf{q}+\mathbf{b}\in\mathbb{R}^{d_{F}}$ according to
\[
c_{r}(\mathbf{\mathbf{q}}) = F(\mathbf{A}\mathbf{q}+\mathbf{b}) \geq0,
\]
where $F:\mathbb{R}^{d_{F}}\mapsto\mathbb{R}$ is some non-linear piecewise smooth function and $\mathbf{A}$ represents a sparse matrix. Within each integrator
step, each of the $d_{F}$ elements of $\mathbf{w}(t)=\mathbf{A}\mathbf{q}(t)+\mathbf{b}=\mathbf{A}(\mathbf{S}\bar{\mathbf{q}}(t)+\mathbf{m})+\mathbf{b}$
are represented as a cubic polynomial.
Then, locating boundary collisions of $\mathbf{w}(t)$ amounts to solving non-linear univariate root finding problems, for which  standard numerical methods can be used.
Note that representing $\mathbf{w}(t)$ as a polynomial means that solving these root-finding problems does not require repeatedly computing any model quantity other than the value of $F$.
Therefore, locating  boundary collisions does not incur large  computational costs as long as the values of $F$ and its gradient are computationally cheap to determine.

\subsection{Illustrative examples of samples from constrained NGRHMC processes\label{subsec:Illustrative-examples}}
\begin{figure}
\begin{centering}
\includegraphics[scale=0.5]{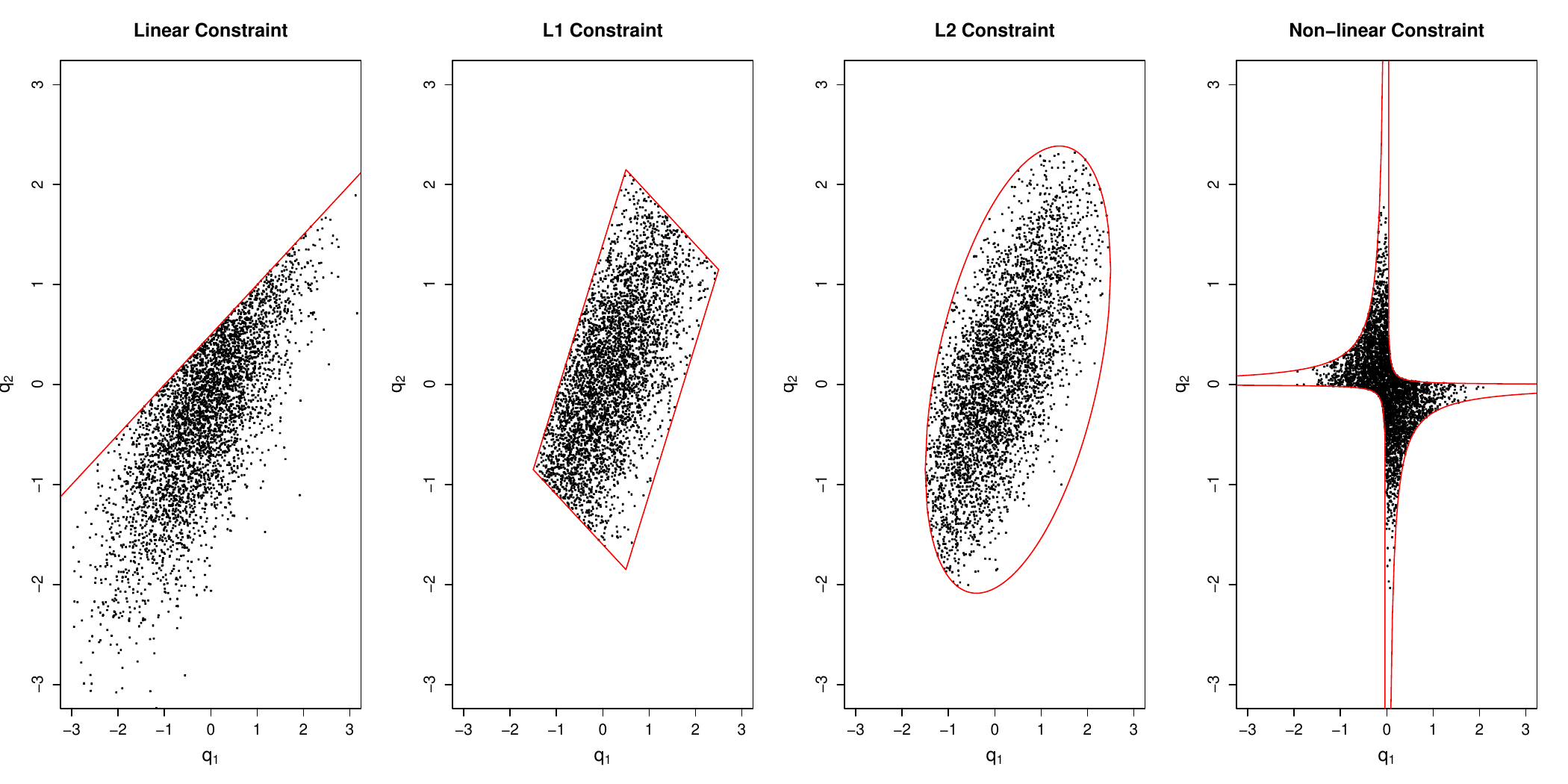}\caption{\label{fig:Discrete-samples-of-illustration}Discrete samples of NGRHMC
processes for a bivariate normal distribution subject to different
types of constraints. See Section \ref{subsec:Illustrative-examples}
for details.}
\par\end{centering}
\end{figure}
\begin{figure}
    \centering
    \includegraphics[scale=0.6]{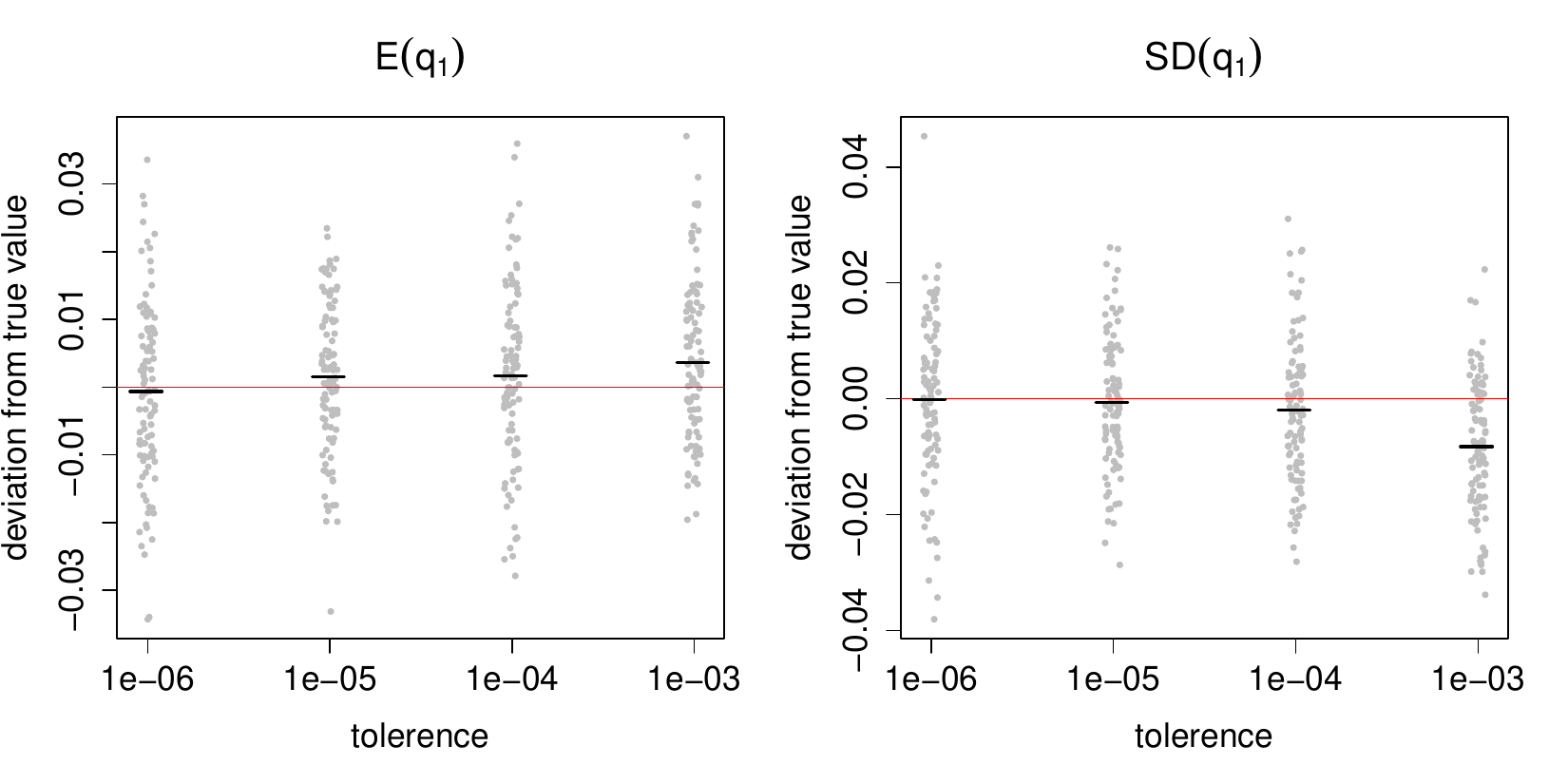}
    \caption{Assessment of bias in estimating $E(q_1)$ (left panel) and $SD(q_1)$ (right panel) as function of the integrator tolerance for the model with linear constraints discussed in Section \ref{subsec:Illustrative-examples}. Grey dots (horizontal coordinate jittered for readability) indicate the deviations of 100 repeated estimates of $E(q_1)$ and $SD(q_1)$ and their true values
     (see latter part of Section \ref{subsec:Illustrative-examples} for details).  The black short lines indicate the bias, calculated as the mean of the 100 repeated estimates.}
    \label{fig:bias-assess}
\end{figure}

In Figure \ref{fig:Discrete-samples-of-illustration} we illustrate the constrained NGRHMC approach. It plots discrete samples from constrained NGRHMC processes for a bivariate normal distribution with a mean of zero, unit marginal variances, and a correlation of 0.75 subject to various constraints.
The leftmost panel shows the result obtained for the constrained NGRHMC process for the bivariate normal subject to the linear constraint $q_{1}-2q_{2}+1\geq0$, for which the implementation described in Section  \ref{subsec:Linear-constraints} is used.
In the two middle panels  we see   NGRHMC samples  for the bivariate normal  subject to the
$\ell_1$- and $\ell_2$-norm constraints on  $\mathbf{w}=[q_{1}-1/2,q_{1}-q_{2}/2+1/10]$ with $v=2$,
which are obtained from using the implementations outlined in Section \ref{subsec:norm-constraints}.
Finally, the rightmost panel shows a  NGRHMC sample  for the non-linear constraint that the spectral radius of the matrix
\[
\left[\begin{array}{cc}
0.8 & q_{2}\\
q_{1} & 0.9
\end{array}\right]
\]
is less than one, using the implementation explained in Section \ref{subsec:Non-lin-Constraints-on-the}.

To assess the extent of the bias in a constrained NGRHMC estimation process caused by using the numerical Hamiltonian trajectories without any bias correction (such as that based on a MH step in a standard HMC), we conducted a further simulation experiment. 
There we considered the estimation of the mean $E(q_1)$ and the standard deviation $SD(q_1)$ for the above bivariate Gaussian distribution subject to the linear constraint. The estimation process was implemented  using different levels of the error tolerance in the BS Runge-Kutta integrator, where for each error tolerance level the estimation process was repeated 100 times. Each NGRHMC estimation process was performed with the default  settings of the pdmphmc package, which produces 4 independent trajectories with $T=10000$, 
where their first halves are used for adaption and then discarded. The remaining part of each trajectory was  sampled 
1000 times at equidistant points in time to produce based on the resulting samples from the 4 trajectories the estimates for $E(q_1)$ and $SD(q_1)$,  which are compared to their true values. The true values  are obtained by using highly precise numerical integration.

The deviations of the 100 repeated estimates of $E(q_1)$ and $SD(q_1)$ from their true values obtained for 4 error tolerance levels ($1e-06$, $1e-0.5$, $1e-04$, $1e-03$)  are represented by dots in the plots of  Figure  \ref{fig:bias-assess}. Also shown is the bias computed from the difference of the mean of the 100 repeated  estimates and the true value. It is seen that for all considered error tolerances, the bias is  small relative to the overall MC variation of the 100 repeated estimates and can be considered to be negligible  for a tolerance level of $1e-04$  (which is default in pdmphmc, and used throughout the rest of this paper) and smaller.

\subsection{Constrained versus unconstrained for transformed variates\label{subsec:Unconstrained-vs-Transfromation}}
\begin{table}
\centering{}%
\begin{tabular}{cccccccccc}
\hline
 & \multicolumn{4}{c}{Unconstrained NGRHMC for } &  & \multicolumn{4}{c}{Constrained NGRHMC}\tabularnewline
 & \multicolumn{4}{c}{transformed parameters} &  & \multicolumn{4}{c}{}\tabularnewline
\cline{2-5} \cline{3-5} \cline{4-5} \cline{5-5} \cline{7-10} \cline{8-10} \cline{9-10} \cline{10-10}
 & ESS & ESS/s & MCSD & MCSD &  & ESS & ESS/s & MCSD & MCSD\tabularnewline
 &  D  &  D    & D & C       &  &  D  &   D   & D & C\tabularnewline
\hline\\[-0.1cm]
 & \multicolumn{9}{c}{iid Gaussian model (\ref{eq:toy-model-1})}\tabularnewline
$\mu$ & 10515 & 6729 & 0.004 & 0.003 &  & 4616 & 2283 & 0.006 & 0.006\tabularnewline
$\sigma$ & 9698 & 6206 & 0.005 & 0.003 &  & 6233 & 3083 & 0.006 & 0.005\tabularnewline\\
 & \multicolumn{9}{c}{AR(1) model (\ref{eq:toy-model-2})}\tabularnewline
$\phi$ & 6005 & 732 & 1.5e-4 & 9.8e-5 &  & 9041 & 1411 & 1.3e-4 & 6.5e-5\tabularnewline
$\sigma$ & 8669 & 1066 & 8.8e-5 & 3.1e-5 &  & 8319 & 1298 & 9.1e-5 & 3.1e-5\tabularnewline\\
 & \multicolumn{9}{c}{Gaussian mixture model (\ref{eq:toy-model-3})}\tabularnewline
$\mu_{1}$ & 4233 & 25 & 0.0029 & 0.0028 &  & 4681 & 72 & 0.0028 & 0.0026\tabularnewline
$\mu_{2}$ & 4510 & 26 & 0.0030 & 0.0027 &  & 4908 & 76 & 0.0028 & 0.0027\tabularnewline
$\sigma$ & 4640 & 27 & 0.0016 & 0.0014 &  & 5076 & 78 & 0.0015 & 0.0014\tabularnewline
\hline
\end{tabular}\caption{\label{tab:Diagnostics-for-toy-models}
Diagnostics for the numerical efficiency of the NGRHMC estimates of the posterior mean for the parameters of the models (\ref{eq:toy-model-1})-(\ref{eq:toy-model-3}).
The  estimates of the posterior means are computed as  continuous-time averages (C) and discrete sample averages (D) from NGRHMC trajectories of time-length $T_{max}=10,000$, where the first half is discarded as burn-in.
The discrete sample averages are computed for  samples of size $N=1000$,  which are taken from the  NGRHMC trajectories after the burn-in period. The MSCD values are the  MC standard deviations of the posterior mean estimates
resulting from 10 independent NGRHMC trajectories. The  ESS and ESS/s statistics  are sample averages from the 10 independent NGRHMC trajectories.}
\end{table}

\begin{figure}
\begin{centering}
\includegraphics[scale=0.5]{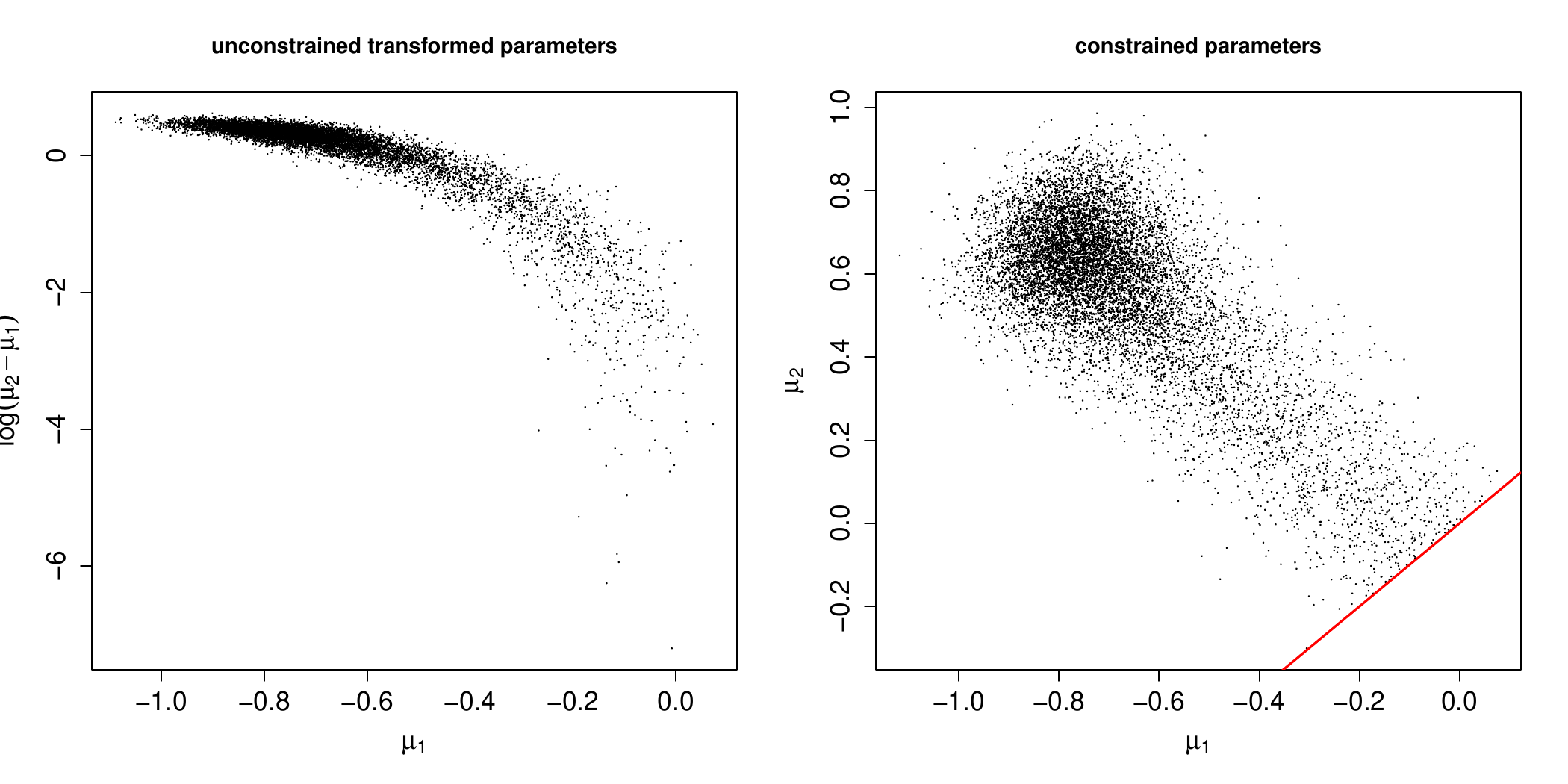}\caption{\label{fig:Discrete-samples-toy-example}
Scatterplots of discrete samples of NGRHMC
processes targeting the posterior distribution for the parameters of the
mixture model (\ref{eq:toy-model-3}); Left panel: Sample from the unconstrained NGRHMC
for the transfomred parameters $(\mu_{1},\log(\mu_{2}-\mu_{1}))$; Right panel: Sample from the constrained NGRHMC
for the original  parameters $(\mu_{1},\mu_{2}$; The red line is the boundary of the constrained parameter domain.}
\par\end{centering}
\end{figure}

In certain cases, constraints can be imposed on a distribution $\pi(\mathbf{q})$ by transforming $\mathbf{q}$ with a bijective mapping such that the resulting target distribution for the transformed $\mathbf{q}$ is unconstrained and can be simulated without having to explicitly account for a constraint.
Examples for such transformations  are a log-  or a logit-transformation.
To analyze the relative merits of the proposed constrained NGRHMC method compared to this parameter transformation approach, we conducted MC simulation experiments in which we consider as target distributions the posterior for the parameters of some simple Bayesian models subject to parameter restrictions.

The first model is a Gaussian model, where
\begin{equation}
y_{i}\sim\text{iid }N(\mu,\sigma^{2}),\quad i=1,\dots,n,\quad\mu>0,\label{eq:toy-model-1}
\end{equation}
with a flat prior for $\mu$ on the interval $(0,\infty)$ and an Exp(1) prior for $\sigma$.
Here, an alternative to using the constrained NGRHMC approach to simulate the posterior of $\mathbf{q}=(\mu,\log(\sigma))^{T}$ with $\mu>0$ is to simulate the corresponding posterior for the transformed parameters $\mathbf{q}^*=(\log(\mu),\log(\sigma))^{T}$ with an unconstrained NGRHMC. The data used in our experiment comparing the two approaches are $(y_1,y_2,y_3,y_4)=(-1,-0.3,0.3,1.2)$.
The second model is a stationary Gaussian first-order autoregression (AR(1)) with positive serial correlation given by
\begin{equation}
y_{t}|y_{t-1}\sim N(\phi y_{t-1},\sigma^{2}),\quad t=2,\dots,n,\quad 0<\phi<1,\label{eq:toy-model-2}
\end{equation}
where we assume a uniform prior for $\phi$ on the interval $(0,1)$ and an Exp(1) prior for $\sigma$. In this case one can simulate the posterior of $\mathbf{q}=(\phi,\log(\sigma))^{T}$ subject to the constraints $\phi>0$ and  $\phi<1$ or the unconstrained posterior of $\mathbf{q}^*=(\log(\phi/(1-\phi)),\log(\sigma))^{T}$. The data we use for the experiment are $n=100$ observations  simulated from the model with parameter values set to $(\phi,\sigma)=(0.99,0.1)$.
The last model is a two-component Gaussian mixture model for an iid variable $y_i$, $i=1,\ldots,n$, with density
\begin{equation}
p(y|\mu_{1},\mu_{2},\sigma)=\frac{1}{2}\mathcal{N}(y|\mu_{1},\sigma^{2})+\frac{1}{2}\mathcal{N}(y|\mu_{2},\sigma^{2}),\quad \mu_{1}\leq\mu_{2},\label{eq:toy-model-3}
\end{equation}
where we set  flat priors for $\mu_1$ and $\mu_2$ on  $(-\infty,\infty)$ and $[\mu_1,\infty)$, respectively, and  an Exp(1) prior for $\sigma$. Instead of  simulating the posterior of $\mathbf{q}=(\mu_1,\mu_2,\log(\sigma))^{T}$ with the constraint $\mu_{1}\leq\mu_{2}$, one can simulate the unconstrained posterior for $\mathbf{q}^*=(\mu_{1},\log(\mu_{2}-\mu_{1}),\log(\sigma))^{T}$. In the experiment for this model, we take $n=200$ observations simulated from the model with parameter values of $(\mu_1,\mu_2,\sigma)=(-0.5,0.5,1)$.

In Table \ref{tab:Diagnostics-for-toy-models}, we provide diagnostic results for the numerical precision of the NGRHMC estimates of the posterior mean for the parameters of the three models given in Equations  (\ref{eq:toy-model-1})-(\ref{eq:toy-model-3}) resulting for the constrained NRGHMC and the unconstrained transformed parameter NGRHMC approach. It reports the MC standard deviation (MCSD), the effective sample size (ESS) \citep{geyer1992} and  the ESS per second of CPU time (ESS/s) for the NGRHMC posterior mean estimates of the parameters. The  ESS, ESS/s and MCSD values for the Gaussian model (\ref{eq:toy-model-1}) all show that imposing the restriction on $\mu$ through a log-transformation and using the unconstrained NGRHMC results in a more efficient sampling  from the posterior than our constrained NGRHMC approach.  For the AR(1) model (\ref{eq:toy-model-2}), we find that the ESS/s clearly favors the constrained NGRHMC over the unconstrained approach based on the logit-transformation of $\phi$, while the MSCD values for both approaches are almost the same. The poorer sampling efficiency  of the transformation approach is likely due to the fact that the unconstrained posterior resulting from the logit-transformation is highly skewed. The ESS/s values for the mixture model (\ref{eq:toy-model-3}) show that in this case  constrained NGRHMC sampling with $\mu_1 \leq \mu_2$ is roughly three times more efficient than unconstrained NGRHMC sampling  based on the transformation
$\log(\mu_{2}-\mu_{1})$. This substantial efficiency loss of the transformation approach  indicates that the parameter transformation leads to a posterior, which is much more difficult to explore by an NGRHMC process than the posterior of the explicitly constrained posterior. To illustrate this, we compare in Figure \ref{fig:Discrete-samples-toy-example} the scatterplot of the discrete unconstrained NGRHMC sample for  $\mu_{1}$ and $\log(\mu_{2}-\mu_{1})$ with that  of the  constrained NGRHMC sample for  $\mu_{1}$ and $\mu_{2}$.
It can be seen that it is the strong non-linear dependence structure introduced by the transformation in the target density that appears to hinder efficient exploration by Hamiltonian dynamics.

These results illustrate that the choice between the constrained NGRHMC and a transformation approach depends very much on the specific application.
In cases where parameter transformations result in significant non-linearity or other deviations from a Gaussian distribution, it seems advisable to avoid such transformations, while for
transformations that result in target distributions that are close to a Gaussian, the transformation approach is in general recommended.
A transformation that always seems advisable is the log-transformation of the variance in Gaussian models to implement the non-negativity restriction, as it leads to constant scale properties \citep[see e.g.~][]{doi:10.1080/10618600.2019.1584901}.
However, certain constraints cannot be readily imposed by an appropriate bijective transformation, such as constraints that involve a larger set of interdependent non-linear restrictions (as considered in our applications discussed in Section \ref{sec:illustrations} below). In this case, methods such as the one we propose are required.

\section{Illustrations\label{sec:illustrations}}
In this section we illustrate the proposed constrained NGRHMC method on four real-word data sets. For all results, the pdmhmc package was used on a 2020 MacBook Pro.

\subsection{Regularized logistic regression\label{subsec:Constrained-Logistic-regression}}

\begin{table}
\begin{centering}
\begin{tabular}{llccccccccccc}
\hline
 && \multicolumn{4}{c}{ $\ell_1$ constraint} &  & \multicolumn{4}{c}{ $\ell_2$ constraint} &  & unconstr.\tabularnewline
\cline{3-6} \cline{4-6} \cline{5-6} \cline{6-6} \cline{8-11} \cline{9-11} \cline{10-11} \cline{11-11} \cline{13-13}
$s$                        && 0.2 & 0.5 & 1.0 & 1.5 &  & 0.2 & 0.5 & 1.0 & 1.5 &  & \tabularnewline
\hline
$\delta$                   &ESS  & 9219 & 8427 & 7221 & 7382 &  & 9107 & 8681 & 7094 & 7720 &  & 7489\tabularnewline
                           &ESS/s& 63 & 47 & 52 & 53 &  & 77 & 67 & 53 & 55 &  & 55\tabularnewline
                           &$\hat R$& 1.0002 & 1.0002 & 1.0008 & 0.9998 &  & 1.0002 & 1.0007 & 1.0003 & 1.0003 &  & 1.0005\tabularnewline\\[-0.2cm]

$\boldsymbol{\beta}$       &min ESS  & 1671 & 2578 & 6888  & 6121  &  & 3152  & 3119  & 7395  & 6580  &  & 6461\tabularnewline
                           &max ESS  & 2301 & 2902 & 8551  & 25874  &  & 4026  & 4676  & 9497  & 23507  &  & 25441\tabularnewline
                           &min ESS/s& 11 & 14 & 50 & 44  &  & 27  & 24  & 55  & 47  &  & 47\tabularnewline
                           &max $\hat R$     & 1.0010 & 1.0008 & 0.9997  & 1.0001 &  & 1.0001  & 1.0008  & 0.9998  & 1.0002  &  & 0.9999\tabularnewline
\hline
\end{tabular}\caption{\label{tab:Effective-sample-sizes-logistic}
Diagnostic results for discrete NGRHMC samples from the posterior of the parameters in the logistic regression model (\ref{eq:logistic regression}) with constraints
(\ref{eq:logistic-constraints}) and without constraints (right-most column). The figures for
$\boldsymbol{\beta}$  are the minimum ESS and ESS/s and the maximum ESS and $\hat R$ value over all parameters in $\boldsymbol{\beta}$.
$\hat R$ is computed from 8 independent NGRHMC trajectories of length $T_{max}=10,000$, where the first half of the trajectories is discarded as burn-in. The ESS and ESS/s statistics are sample averages computed from these 8 independent NGRHMC trajectories.}
\par\end{centering}
\end{table}

\begin{figure}
\begin{centering}
\includegraphics[scale=0.5]{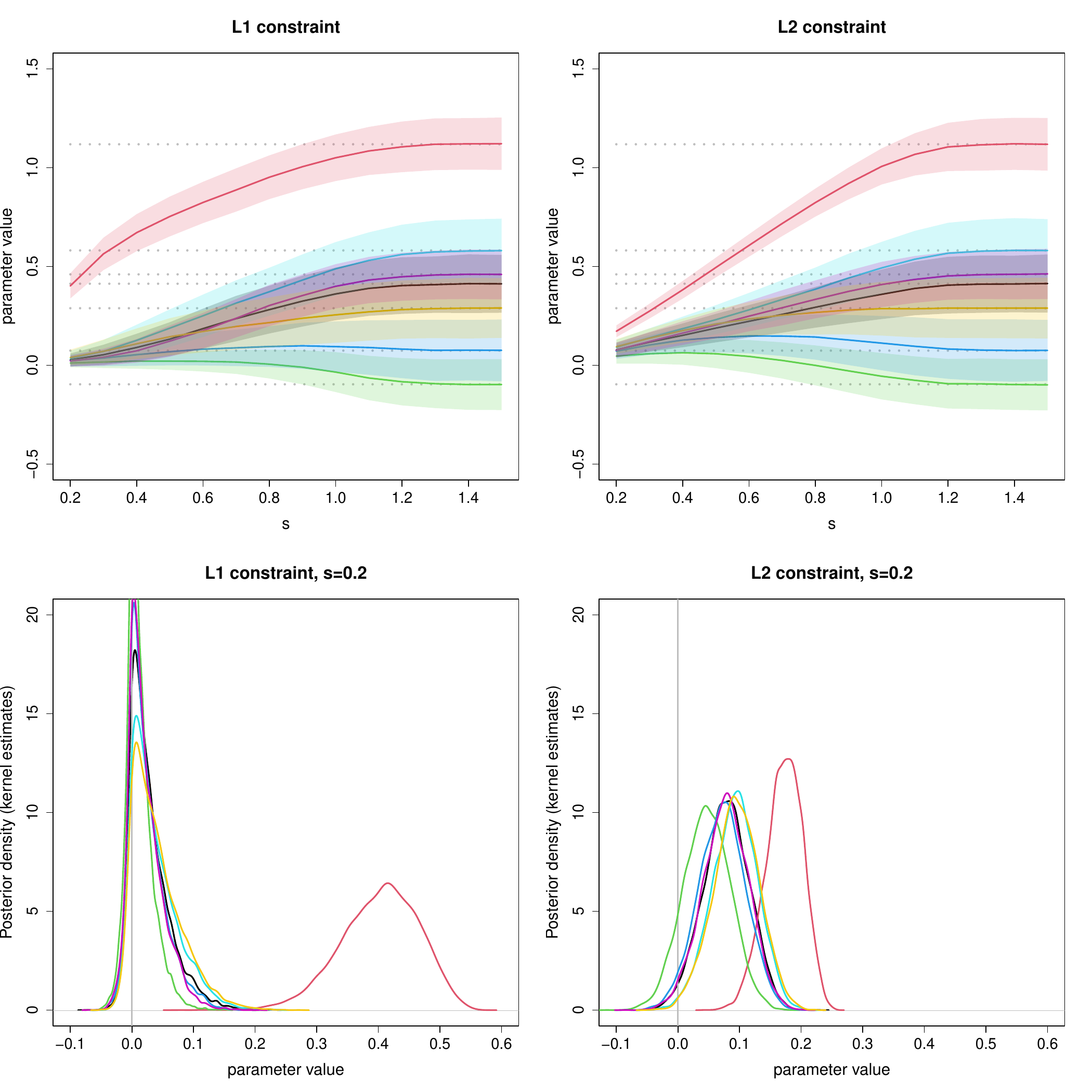}\caption{\label{fig:The-parameter-vector-logistic}
Posterior of the $\boldsymbol{\beta}$ parameters in the logistic regression model (\ref{eq:logistic regression}) with constraints
(\ref{eq:logistic-constraints}) for different values of the regularizing coefficient  $s$ (upper panels), and posterior density estimates for the $\boldsymbol{\beta}$ parameters for $s=0.2$ (bottom panels).
The solid lines in the upper panels
indicate the posterior mean of each element of $\boldsymbol{\beta}$,
and the shaded regions are the corresponding posterior mean $\pm$
1 posterior standard deviation. Dotted lines indicate posterior means
without constraints. The statistics plotted are sample averages calculated from 8 independent NGRHMC trajectories of length $T_{max}=10,000$, where the first half of the trajectories is discarded as burn-in period.}
\par\end{centering}
\end{figure}

In the first illustrative application, we use a regularized Bayesian logistic regression model on a subset of the Pima data, which include outcomes for diabetes tests performed on women of Pima Indian heritage and several predictors \citep{ripley86}.
It has previously been analyzed using a regularized Bayesian logistic regression by \cite{gram:pols:2012}, and a PDMP application to  inference in a Bayesian logistic regression with inequality constraints can be found in \cite{BIERKENS2018148}.

The logistic  regression model for the observed binary outcome variable $y_i$, $i=1,\dots,532$ is given by
\begin{equation}
\text{logit}(P\{y_{i}=1|\mathbf{x}_{i}\})=\delta+\mathbf{x}_{i}^{T}\boldsymbol{\beta}\label{eq:logistic regression}
\end{equation}
where $\mathbf{x}_{i}\in\mathbb{R}^{7}$ are the standardized predictors. The priors assumed for the regression coefficients $\delta$ and $\boldsymbol{\beta}$ are
$\delta\sim N(0,10^{2})$ and $\boldsymbol{\beta}\sim N(\boldsymbol{0}_7,10^{2}\mathbf{I}_7)$ combined with either an $\ell_1$- or an $\ell_2$-norm constraint of the following form:
\begin{equation}
\parallel\boldsymbol{\beta}\parallel_{1}\leq s\parallel\hat{\boldsymbol{\beta}}\parallel_{1},\;\text{ or }\;\parallel\boldsymbol{\beta}\parallel_{2}\leq s\parallel\hat{\boldsymbol{\beta}}\parallel_{2},\quad s>0\label{eq:logistic-constraints},
\end{equation}
where $\hat{\boldsymbol{\beta}}$ denotes the unconstraint maximum likelihood (ML) estimate for $\boldsymbol{\beta}$. The smaller the regularizing coefficient $s$, the more the posterior mean of $\boldsymbol{\beta}$ is shrunk towards the zero vector,
similar to a classical Lasso (Least absolute shrinkage and selection operator) and Ridge regression estimator \citep{Hastie_elements}. (The unconstrained ML estimate used to define the bounds serves as a benchmark to normalize $s$ so that values for $s$ are comparable across different data sets.)
The specific implementations of the constrained NGRHMC method used for the logistic regression with the $\ell_1$- and $\ell_2$-restriction are described in   Section \ref{subsec:norm-constraints}.

In the upper panels of Figure \ref{fig:The-parameter-vector-logistic} we plot the paths of the mean for the $\boldsymbol{\beta}$ parameters  together with the corresponding 2-standard-deviation  intervals  under the constrained posterior  for decreasing values of $s$, and in the bottom panels the posterior density estimates for the $\boldsymbol{\beta}$ parameters for $s=0.2$.
It is seen that for the tightest bound considered ($s=0.2$), the center of the $\ell_1$-constrained posteriors are closer to zero for all parameters than those under the $\ell_2$ constraint, except for one parameter.
This mirrors the fact that the frequentistic  Lasso estimator with $\ell_1$ regularization typically results in sparse parameterizations of the fitted model, whereas the  $\ell_2$-regularized Ridge estimator does not \citep{Hastie_elements}.

Table \ref{tab:Effective-sample-sizes-logistic} reports  the ESS and ESS/s  for the NGRHMC samples from the posterior of the model parameters  for selected values of $s$.
Also reported is the potential scale reduction factor $\hat R$, which is used to check convergence  of the  NGRHMC samples \citep{gelm:rubin:1992,brook:gelm:1998}. For $\hat R<1.01$ it can be assumed that convergence has been achieved \citep{vehtari:etal:2021}.
Since the $\hat R$ values are all well below this threshold, the NGRHMC samples appear to have converged to the constrained posterior distribution.
The ESS and ESS/s values show that  the sampling efficiency of the constrained NGRHMC algorithm deteriorates as $s$ decreases compared to the unconstrained case.
However, in this application the necessary computing times for a reliable Bayesian inference based on the constrained NGRHMC are by no means long:
For example, the result that for $s=0.2$ the minimum ESS/s over all parameters under the $\ell_1$-constraint is 11 shows that approximately only 15 minutes are required to achieve a sampling efficiency as for 10,000 hypothetical i.i.d.~draws from the joint posterior of all parameters.

\subsection{Identification restrictions for neural nets}

\begin{figure}
\begin{centering}
\includegraphics[scale=0.5]{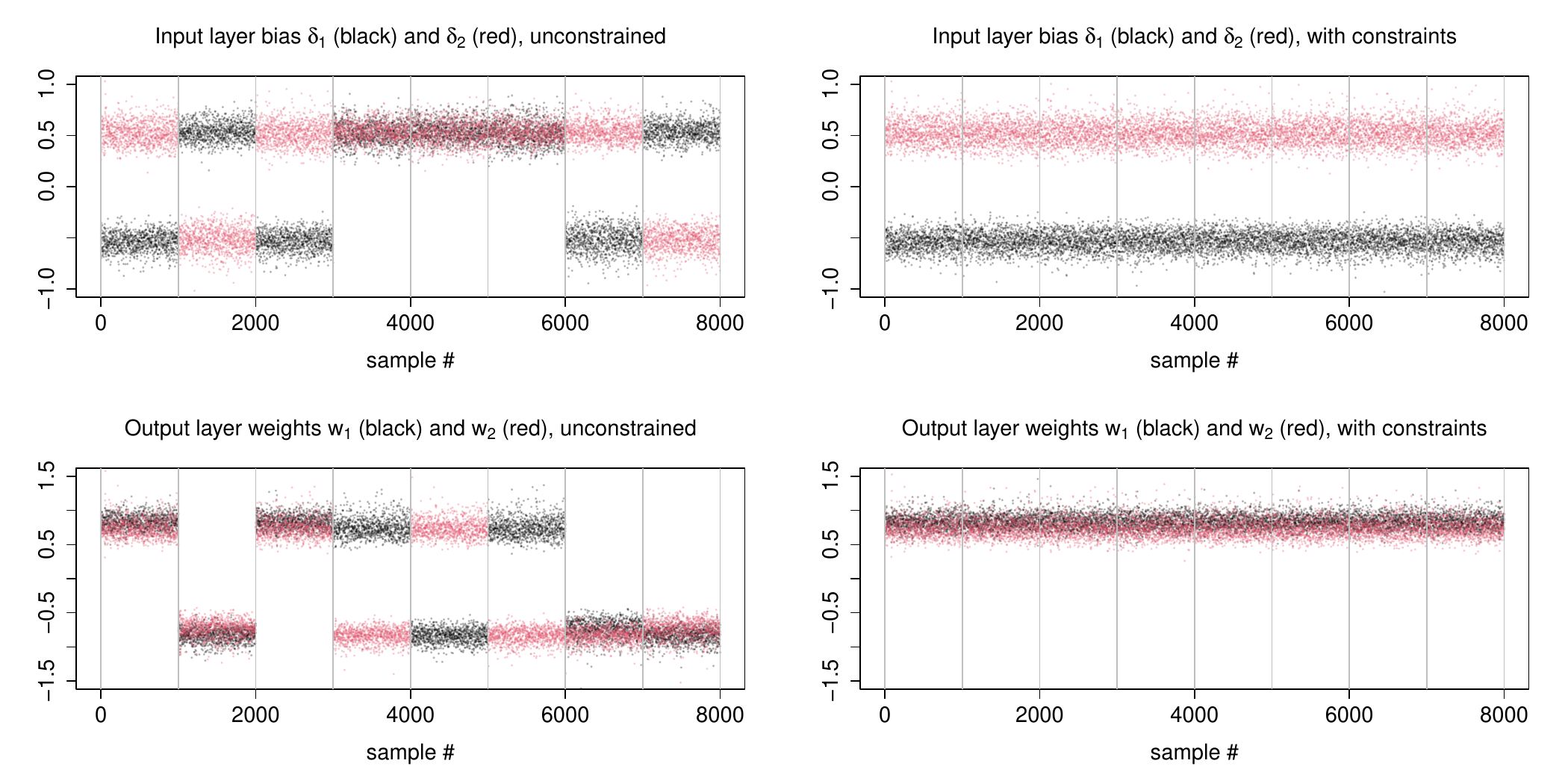}
\caption{\label{fig:NGRHMC-output-for-neural}
Trace plots of 8 independent discrete  NGRHMC samples from the posterior of the parameters $(\delta_1, \delta_2)$ and $(w_1, w_2)$ in the
neural network model  (\ref{eq:neural_network_model}) applied to the simulated data set. Left panels:  NGRHMC samples without the constraints (\ref{eq:network-constraints}); Rights panels: NGRHMC samples with the constraints (\ref{eq:network-constraints}).
The trace plots of the 8 independent NGRHMC samples, each with a size of $N=1000$, are separated by the gray vertical lines. The true parameter values are  $\delta_1=0.5$, $\delta_2=-0.5$ and $w_1=w_2=1$.}
\par\end{centering}
\end{figure}

In the second application, we consider a Bayesian neural network model \citep{9756596}.
Without any normalizing restrictions, the parameters in such a model are not statistically identified (in the classical sense) as it has many observationally equivalent parameterizations, which typically leads to complicated multimodal posterior distributions. There are several strategies to deal with this issue \citep[][Section IV-C-2]{9756596}.
One of them, which can  be straightforwardly implemented with the constrained NGRHMC approach, is to remove unidentifiability by imposing  normalizing restrictions on the parameters through the prior.

The model considered here  is a regression-type neural network with a single hidden layer consisting of $J$ neurons. It has the form
\begin{equation}
y_{i}\sim N(\mu_{i},\sigma^{2}),\quad\; \mu_{i}=\alpha+\sum_{j=1}^{J}w_{j}g(\eta_{j,i}),\quad\; \eta_{j,i}=\delta_{j}+\mathbf{x}_{i}^{T}\boldsymbol{\beta}_{j},\label{eq:neural_network_model}
\end{equation}
where $y_i$, $i=1,\ldots,n$ is the outcome variable, $\mathbf{x}_{i}\in\mathbb{R}^{p}$ is a vector of covariates, and $g$ represents the activation function, which is specified by $g(x)=-1+2\exp(x)/(1+\exp(x))$.
The model parameters are $\boldsymbol{\theta}=(\alpha,\{w_{j}\}_{j=1}^{J},\{\delta_{j}\}_{j=1}^{J},\{\boldsymbol{\beta}_{j}\}_{j=1}^{J})$   and $\sigma$.
Since we apply the model to standardized data, we use independent standard normal priors for all parameters in $\boldsymbol{\theta}$ and an Exp(1) prior for $\sigma$.

In model (\ref{eq:neural_network_model}),
unidentifiability results from a `scaling symmetry' and a `weight-space symmetry' \citep{9756596}.
The scaling symmetry means that due to the equality $g(x)= -g(-x)$, the $j$th neuron  makes the same contribution to the prediction  $\mu_i$ for   $(w_{j},\delta_{j},\boldsymbol{\beta}_{j})$ and $(-w_{j},-\delta_{j},-\boldsymbol{\beta}_{j})$, regardless of the value for $\mathbf{x}_{i}$.
The weight-space symmetry implies that for $J\geq2$, swapping the values for $(w_{j},\delta_{j},\boldsymbol{\beta}_{j})$ and $(w_{k},\delta_{k},\boldsymbol{\beta}_{k}),\;k\neq j$ does not change the value for $\mu_i$ either.
These symmetries can be removed by combining the assumed standard normal priors for the $w_j$  and the $\delta_j$ parameters with the  restrictions,
\begin{equation}
w_{j}\geq0,\;j=1,\dots,J,\quad   \text{and}\quad \delta_{j}\geq\delta_{j-1},\;j=2,\dots,J, \label{eq:network-constraints}
\end{equation}
where the first set of restrictions eliminate the scaling symmetry and the second set the weight-space symmetry.
These restrictions have a linear form for which the implementation of the constrained NGRHMC is described in Section \ref{subsec:Linear-constraints}.

We apply the model to an artificial  and a real-world data set. The artificial data are generated  by simulating the model as given in Equation (\ref{eq:neural_network_model}) using a similar design as for the `Sum of sigmoids' data in \citet[Section 11.6]{Hastie_elements}. For the simulated model, $J=2$ and $p=2$. The parameter values are set to $\alpha=0$, $\delta_{1}=0.5$, $\delta_{2}=-0.5$,
$\boldsymbol{\beta}_{1}=(1,0)^{T}$, $\boldsymbol{\beta}_{2}=(-0.1,1)^{T}$, $w_{1}=w_{2}=1$, $\sigma=0.1$ and the sample size is $n=200.$ The real-world dataset comes from the study by  \citet{STAMEY19891076} on the determinants of the prostate-specific antigen level and is also used in \citet{Hastie_elements}.
It contains the measured values for the antigen level in 97 men and the values of 8 predictors.

In Figure \ref{fig:NGRHMC-output-for-neural} we compare NGRHMC samples from the posterior of the parameters $(\delta_1, \delta_2, w_1, w_2)$
with and without the normalizing restrictions (\ref{eq:network-constraints}) for the model applied to the simulated data.
It shows the trace plots  of 8 independent discrete NGRHMC samples with  a size of $N=1000$ each and  different randomly generated initial values of the NGRHMC process. In the  trace plots for the unconstrained posterior, their multimodal shape is clearly visible.
Depending on the respective initial value of the simulated NGRHMC process, it is apparently attracted to one of the modes and gets stuck in its vicinity throughout the runtime. This indicates that  the regions of the different modes are virtually separated from each other in the sense that the probability for the  NGRHMC process to move from the region of one mode to that of another mode is almost zero.
This prevents the NGRHMC process from exploring the complete unconstrained posterior and makes it nearly impossible for any  HMC-type method to reliably approximate it.
In contrast, the trace plots for the constrained posterior, with the multimodality removed by the normalizing restrictions, show that the constrained NGRHMC algorithm produced well-mixing samples.
This is confirmed by the diagnostic results presented in Table \ref{tab:Diagnostics-results-for-neural} for the NGRHMC samples of the parameters in the normalized model with different numbers of neurons $J$ for the applications to the simulated data set and the prostate cancer data.
We also note that the posterior mean for the standard deviation of the noise component $\sigma$ given in Table \ref{tab:Diagnostics-results-for-neural} - as expected - decreases with increasing $J$. In the application to the simulated data, it decreases most when $J$ is increased from 1 to its true value, $J=2$, but decreases only slightly when $J$ is further increased.
This shows that the estimates for $\sigma$ are  useful for identifying $J$. In the application to the  prostate cancer data, this  suggests choosing $J=2$ since further increases in $J$ result in only small decreases in the $\sigma$ estimate compared to the decrease we find when moving from $J=1$ to $J=2$.

\begin{table}
\centering{}%
\begin{tabular}{llccccccccc}
\hline
 && \multicolumn{4}{c}{Simulated data} &  & \multicolumn{4}{c}{Prostate cancer data}\tabularnewline
\cline{3-6} \cline{4-6} \cline{5-6} \cline{6-6} \cline{8-11} \cline{9-11} \cline{10-11} \cline{11-11}
 && $J=1$ & $J=2$ & $J=4$ & $J=8$ &  & $J=1$ & $J=2$ & $J=4$ & $J=8$\tabularnewline
\hline
$\sigma$ &posterior mean & 0.1204 & 0.0973 & 0.0970 & 0.0968 &  & 0.638 & 0.613 & 0.598 & 0.586\tabularnewline
         &ESS    & 6825 & 5615 & 6259 & 6040 &  & 4862 & 4580 & 4903 & 4961\tabularnewline
         &ESS/s  & 30.0 & 11.5 & 1.2 & 0.5 &  & 57.4 & 18.4 & 7.8 & 3.2\tabularnewline
         &$\hat R$ & 1.0012 & 1.0013 & 1.0012 & 1.0012 &  & 1.0008 & 1.0006 & 1.0005 & 1.0013\tabularnewline\\[-0.2cm]
$\boldsymbol{\theta}$ &min ESS & 10552 & 7790  & 1969 & 3139 &  & 3415 & 5048 & 5559 & 3971\tabularnewline
                      &max ESS & 15430 & 21607  & 9623 & 12189 &  & 15867 & 14018 & 14256 & 11610\tabularnewline
                      &min ESS/s& 46.4 & 16.0 & 0.4 & 0.3 &  & 40.3 & 20.3 & 8.9 & 2.5\tabularnewline
                      &max $\hat R$& 1.0012 & 1.0007 & 1.0031 & 1.0024 &  & 1.0018 & 1.0015 & 1.0013 & 1.0017\tabularnewline
\hline
\end{tabular}\caption{\label{tab:Diagnostics-results-for-neural}
 Diagnostic results for discrete NGRHMC samples from the posterior of the parameters in
 the neural network model  (\ref{eq:neural_network_model})  with constraints
(\ref{eq:network-constraints}) applied to the simulated data set and the prostate cancer data. For the simulated data, the true value of $\sigma$ is 0.1. The figures for
$\boldsymbol{\theta}$  are the minimum ESS and ESS/s and the maximum ESS and $\hat R$ value over all parameters in $\boldsymbol{\theta}$.
$\hat R$ is computed from 8 independent NGRHMC trajectories of length $T_{max}=10,000$, where the first half of the trajectories is discarded as burn-in. All other reported statistics are sample averages computed from these 8 independent NGRHMC trajectories.}
\end{table}

\subsection{Gaussian observations under total positivity constraints}

\begin{table}
\centering{}%
\begin{tabular}{lccccccccccc}
\hline
 & \multicolumn{3}{c}{ESS} & min    & max       &  & \multicolumn{3}{c}{ESS} & min   & max\tabularnewline
 \cline{2-4} \cline{8-10}
 & min & median & max &      ESS/s  & $\hat{R}$ &  & min & median & max      & ESS/s &  $\hat{R}$\tabularnewline
\hline
 & \multicolumn{5}{c}{$\text{MTP}_2$ constrained} &  & \multicolumn{5}{c}{unconstrained}\tabularnewline
\cline{2-6} \cline{3-6} \cline{4-6} \cline{5-6} \cline{6-6} \cline{8-12} \cline{9-12} \cline{10-12} \cline{11-12} \cline{12-12}
$\mathbf{P}_{i,i},\; i=1,\ldots,m$ & 680 & 1795  & 6077  & 0.15  & 1.0072 &  & 5827  & 6908 & 8648  & 2.99  & 1.0028\tabularnewline
$\mathbf{P}_{i,j},\; i\neq j$  & 1292  & 4051  & 7815  & 0.28  & 1.0050 &  & 6502  & 10540  & 24297  & 3.34  & 1.0021\tabularnewline\\[-0.2cm]
$\boldsymbol{\omega}$ & 2890 & 5291  & 6462  & 0.63  & 1.0020  & & 3414 & 9010 & 26632 & 1.75 & 1.0025\tabularnewline
\hline
\end{tabular}\caption{\label{tab:Diagnostics-information-for-MTP2}
Diagnostics results for discrete NGRHMC samples from the unconstrained and $\text{MTP}_2$-constrained posterior of the precision matrix $\mathbf{P}$ in  model (\ref{eq:MTP2-model}) and the GMVP weights in Equation (\ref{eq:GMVP-weights}).
$\hat R$ is computed from 8 independent NGRHMC trajectories of length $T_{max}=10,000$, where the first half of the trajectories is discarded as burn-in. All other reported statistics are sample averages computed from these 8
independent NGRHMC trajectories.}
\end{table}

We now consider estimating the covariance matrix $\boldsymbol{\Sigma}$ for the returns of $m$ assets with an application to the selection of the global minimum  variance portfolio (GMVP) \citep{markowitz:1952}.
If $m$ is large relative to the number of periods of historical return data $n$ used for estimating $\boldsymbol{\Sigma}$, the standard ML estimator is known to be very imprecise. This in turn leads to a poor performance of the resulting estimates and forecasts for the weights
of the GMVP  \citep{ledoit:wolf:2003}. In a recent contribution, \citet{10.1093/jjfinec/nbaa018} propose  to tackle   this estimation
problem by imposing the constraint that the joint distribution of the returns be multivariate totally positive of order 2 ($\text{MTP}_2$). If the joint distribution is Gaussian, the $\text{MTP}_2$ constraint is equivalent to  the restriction that all diagonal elements of the precision matrix $\mathbf{P}=\boldsymbol{\Sigma}^{-1}$ are strictly positive and all non-diagonal ones  non-positive. Thereby it implicitly regularizes the estimation of $\boldsymbol{\Sigma}$ and leads
to sparsity in the  constrained ML estimates of $\mathbf{P}$ \citep{10.1214/17-AOS1668}.

In our application, we consider a Bayesian Gaussian model for the vector of daily  log-returns for $m$ stocks $\mathbf{y}_{t}\in\mathbb{R}^{m}$ of the form
\begin{equation}
\mathbf{y}_{t}\sim\text{ iid }N(\mathbf{0}_m,\mathbf{P}^{-1}),\quad t=1,\dots,n,\label{eq:MTP2-model}
\end{equation}
and set a Wishart prior for $\mathbf{P}$ with  mean  $E(\mathbf{P})=\mathbf{I}_m$ and $m+10$ degrees of freedom.
To enforce the symmetry and positive definiteness of $\mathbf{P}$, it is parameterized by the vector $\mathbf{z}=(z_1,\ldots,z_m,z_{m+1},\ldots,z_{m(m+1)/2})^{T}\in\mathbb{R}^{m(m+1)/2}$ in the decomposition
\begin{equation}
\mathbf{P}=\mathcal{P}(\mathbf{z})=\mathbf{L}(\mathbf{\mathbf{z}})\boldsymbol{\Lambda}(\mathbf{z})\left[\mathbf{L}(\mathbf{z})\right]^{T},\qquad \boldsymbol{\Lambda}(\mathbf{z})=\text{diag}(\exp\{z_{1}\},\dots,\exp\{z_{m}\}),\label{eq:MTP2-model-reparametrized}
\end{equation}
where $\mathbf{L}(\mathbf{z})$ is  a lower triangular matrix with unit diagonal elements  and strictly lower triangular  elements \linebreak $z_{m+1},\ldots,z_{m(m+1)/2}$.
To account for the $\text{MTP}_2$ constraint, the prior  for $\mathbf{z}$ resulting from the assumed Wishart prior for $\mathcal{P}(\mathbf{z})$ is combined with the  restriction
\begin{equation}
\max_{(i,j)\in[1,\dots,m],\;j<i}\left[\mathcal{P}(\mathbf{z})\right]_{i,j}\leq0.\label{eq:MTP2-constraint}
\end{equation}
The prior of $\mathbf{z}$ for a Wishart prior on $\mathcal{P}(\mathbf{z})$ can be found in \cite{2211.01746}. For the constraint NGRHMC applied to this non-linear restriction, we use the specific implementation elements described in   Section \ref{subsec:Non-lin-Constraints-on-the}.
Having estimated $\mathbf{P}$, the estimates for the weights of the GMVP  $\boldsymbol{\omega}\in\mathbb{R}^{m}$ are calculated using the standard GMVP formula given by
\begin{equation}
\boldsymbol{\omega}=\frac{\mathbf{P}\mathbf{1}_{m}}{\mathbf{1}_{m}^{T}\mathbf{P}\mathbf{1}_{m}},\qquad \mathbf{1}_{m}=(1,1,\dots,1)^{T}.\label{eq:GMVP-weights}
\end{equation}

\begin{figure}
\centering{}\includegraphics[scale=0.5]{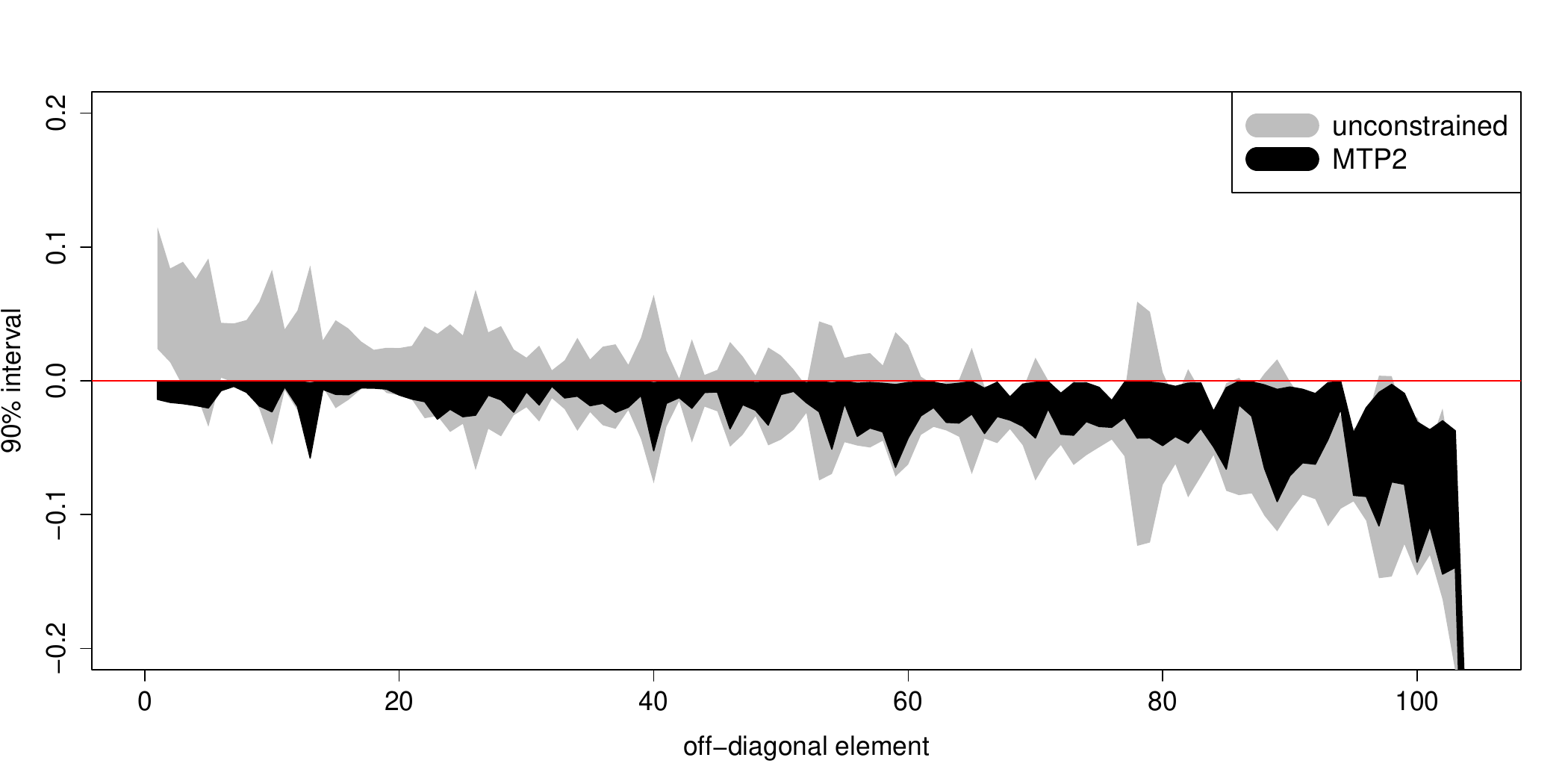}\caption{\label{fig:90=000025-posterior-intervals-MTP2}
NGRHMC approximation to the 90\% credible inetrvals of the 105 different off-diagonal elements of the  precision matrix $\mathbf{P}$ in  model (\ref{eq:MTP2-model}) for the unconstrained and $\text{MTP}_2$-constrained posterior.
The intervals are sorted from left to right by the magnitude of the mean of the unconstrained posterior.}
\end{figure}

\begin{figure}
\begin{centering}
\includegraphics[scale=0.5]{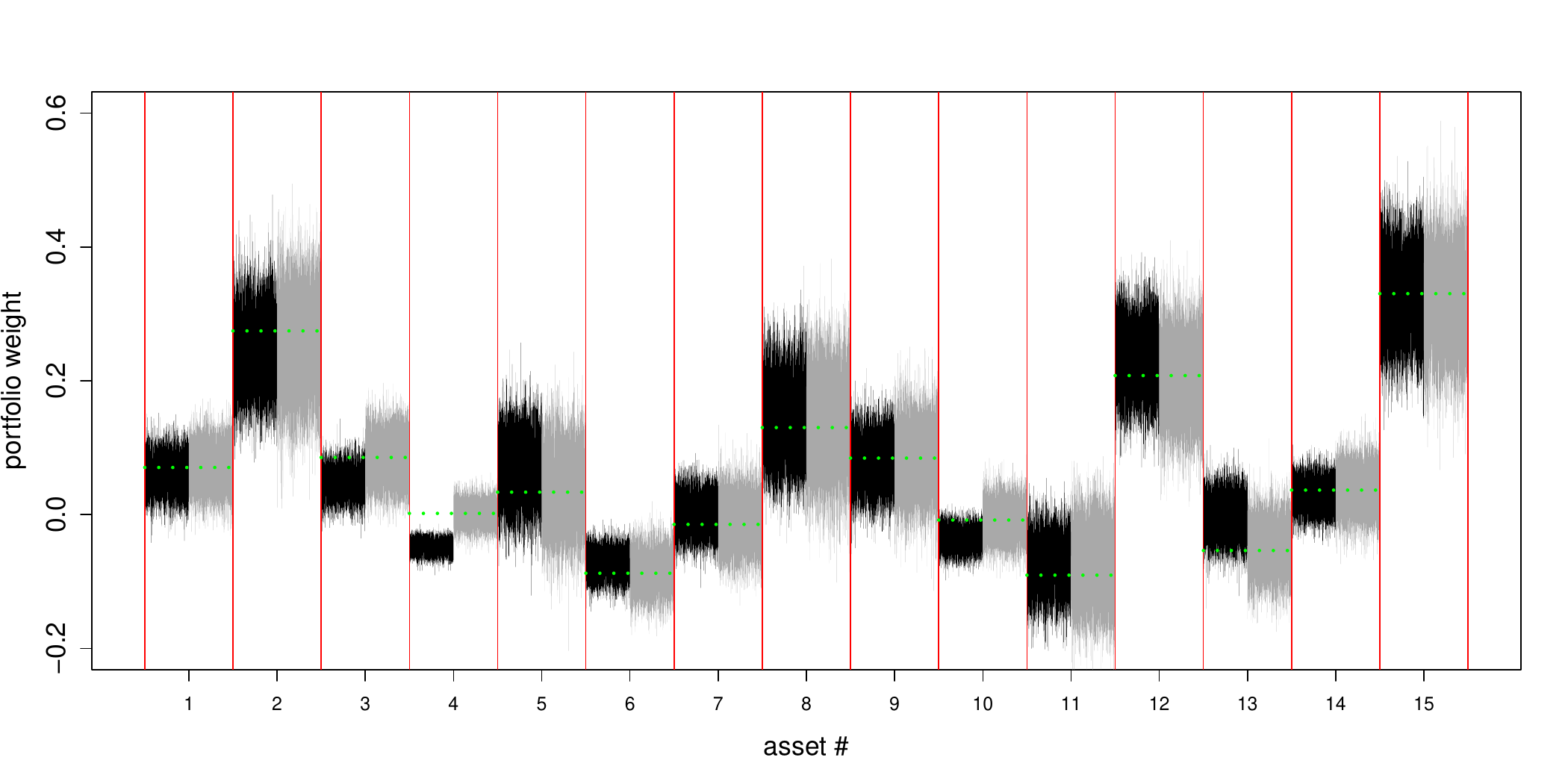}
\par\end{centering}
\caption{\label{fig:Discrete-samples-of-MTP2}
Trace plots of discrete  NGRHMC samples from the posterior of the $m=15$ GMVP weights in $\boldsymbol{\omega}$ defined by Equation (\ref{eq:GMVP-weights}). The black lines are the samples  for the  $\text{MTP}_2$-constrained precision matrix $\mathbf{P}$ and the grey lines the samples for the precision matrix  without the  $\text{MTP}_2$ constraint. The green line are the GMVP weights obtained from the ML estimate of $\mathbf{P}$.}
\end{figure}

We use $n=250$ historical daily returns for $m=15$ stocks  from 02.01.2002 until 27.12.2002. The stocks are randomly selected  from the dataset analyzed by \cite{reh:krueger:liesenfeld:2023}, which consists of  the prices  of all NYSE, AMEX and NASDAQ stocks. Table  \ref{tab:Diagnostics-information-for-MTP2} summarizes the diagnostic results for the NGRHMC  samples from the $\text{MTP}_2$-constrained and unconstrained posterior of $\mathbf{P}$.
The $\hat R$ and ESS/s statistics for the $\mathbf{P}$ elements show that the NGRHMC samples for the unconstrained and constrained posterior have converged, but that the sampling efficiency of the NGRHMC for the constrained posterior is about an order of magnitude lower than for the unconstrained one.
However, this efficiency loss is largely due to the fact that the $\text{MTP}_2$ constraint  changes the shape of the posterior significantly, in a way that makes it not particularly well suited to HMC-type sampling methods. In fact, the marginal constrained posterior is extremely  skewed for some off-diagonal elements of $\mathbf{P}$, while it is almost Gaussian for their unrestricted counterparts.
To illustrate the effects of imposing the  $\text{MTP}_2$ constraint, we provide in Figure \ref{fig:90=000025-posterior-intervals-MTP2} the NGRHMC approximations of the 90\% credible  intervals for the 105 different off-diagonal elements of $\mathbf{P}$.
It can be seen that, as expected, the constraint shifts the marginal posteriors towards the origin and significantly reduces the posterior variances compared to the unconstrained case. This mirrors the result in the classical ML context that the $\text{MTP}_2$ constraint typically leads to sparsity  in the estimated  $\mathbf{P}$  \citep{10.1214/17-AOS1668}.
It is worth noting that the Bayesian shrinkage of $\mathbf{P}$ using the constrained NGRHMC for the parametrization (\ref{eq:MTP2-model-reparametrized}) ensures that the simulated posterior draws of $\mathbf{P}$ are positive definite. Simply shifting the off-diagonal elements of $\mathbf{P}$ towards zero in some arbitrary way easily violates the positive definiteness.

The consequences  of the $\text{MTP}_2$ constraint for the estimates of the  GMVP weights $\boldsymbol{\omega}$ are illustrated in Figure \ref{fig:Discrete-samples-of-MTP2}. It shows the trace plots of the NGRHMC posterior samples of the GMVP weights both with and without the $\text{MTP}_2$ constraint together with  the weights' ML point estimates.  We see that, consistent with the corresponding reduction of the posterior variance of the precision matrix $\mathbf{P}$, the posterior variance of the $\text{MTP}_2$-constrained weights is uniformly smaller than that of their unconstrained counterparts. It is also seen that the  average level of the unconstrained weights  is close to the corresponding ML value, while  in  about half of the cases the level of the $\text{MTP}_2$-constrained weights differs significantly.  The diagnostic results for the constrained NGRHMC samples of the GMVP weights reported in Table 3 indicate a satisfactory simulation efficiency as revealed by  their minimum ESS/s value which is not much less than half of that for the NGRHMC unconstrained samples.

\subsection{Stationarity-constrained VAR model}

\begin{table}
\centering{}%
\begin{tabular}{lccccccccccccc}
\hline
 & \multicolumn{3}{c}{ESS} &  & min ESS/s & max $\hat{R}$ &  & \multicolumn{3}{c}{ESS} &  & min ESS/s & max $\hat{R}$\tabularnewline
\cline{2-4} \cline{3-4} \cline{4-4} \cline{9-11} \cline{10-11} \cline{11-11}
 & min & median & max &  &  &  &  & min & median & max &  &  & \tabularnewline
\hline
 & \multicolumn{6}{c}{with stationarity constraint} &  & \multicolumn{6}{c}{without stationarity constraint}\tabularnewline
\cline{2-7} \cline{3-7} \cline{4-7} \cline{5-7} \cline{6-7} \cline{7-7} \cline{9-14} \cline{10-14} \cline{11-14} \cline{12-14} \cline{13-14} \cline{14-14}
$\mathbf{B}_{i,j}$ & 4926 & 5421  & 5810 &  & 2.6 & 1.0016 &  & 7042  & 10483  & 12661  &  & 3.8  & 1.0025\tabularnewline
$\boldsymbol{\alpha}_{i}$ & 5400 & 5527  & 5878  &  & 2.8  & 1.0009 &  & 7881 & 10707  & 12133 &  & 4.2 & 1.0025\tabularnewline
$\mathbf{P}_{i,j}$ & 5801  & 6097 & 6401 &  & 3.0  & 1.0033 &  & 5902  & 6135  & 6816  &  & 3.1  & 1.0022\tabularnewline
\hline
\end{tabular}\caption{\label{tab:Diagnostics-information-for-VAR} Diagnostic results for discrete NGRHMC samples from the posterior of the parameters in
 the VAR model  (\ref{eq:VAR-model})  with and without the stationarity constraint (\ref{eq:VAR-stationary}).
 $\hat R$ is computed from 8 independent NGRHMC trajectories of length $T_{max}=10,000$, where the first half of the trajectories is discarded as burn-in. All other reported statistics are sample averages computed from these 8
independent NGRHMC trajectories.}
\end{table}

\begin{figure}
\begin{centering}
\includegraphics[scale=0.45]{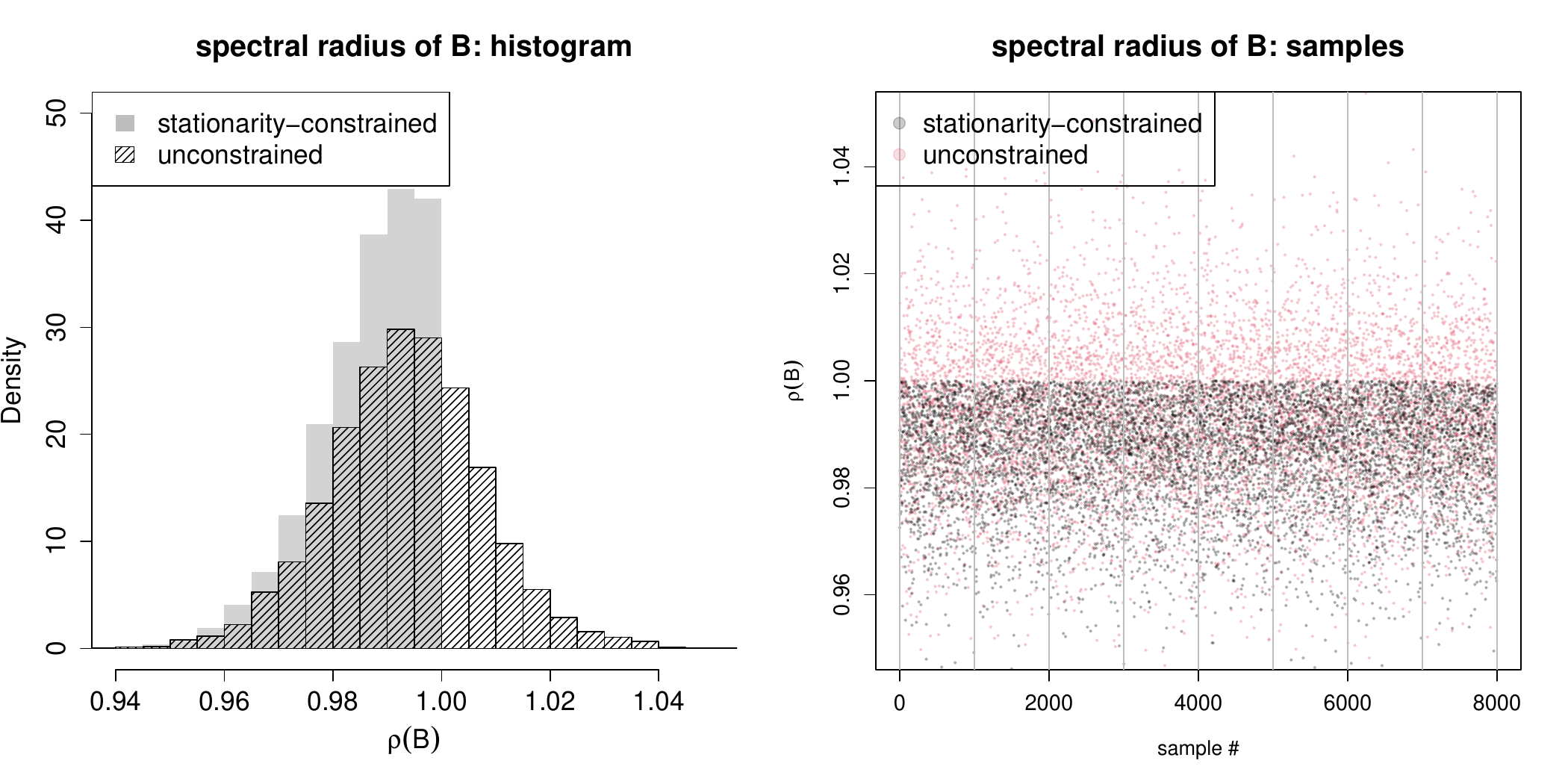}
\caption{\label{fig:VAR_rho_histogram}
Histogram of   discrete NGRHMC samples from the posterior of the spectral radius  $\rho(\mathbf{B})$ for the VAR model (\ref{eq:VAR-model})  with and without the stationarity constraint (\ref{eq:VAR-stationary}) (left panel), and  trace plots of 8 independent discrete NGRHMC samples from the posterior of $\rho(\mathbf{B})$ (right panel). The trace plots of the 8 independent NGRHMC samples, each with a size of $N=1000$, are separated by the gray vertical lines.
}
\par\end{centering}
\end{figure}

We now turn to a Bayesian vector autoregressive (VAR) model with an application to US government bond yields.
VAR models are successfully used for forecasting multivariate time series and analysing the dynamic interaction between time series,
and an application of a  Bayesian VAR  to bond yields can be found, for example, in \cite{CARRIERO20122026}.
In our application we take into account the economically motivated prior information that financial theory is generally based on the assumption that interest rates are stationary \citep{sarno:thornton:valente:2007}.

The VAR model that we consider for the time series of bond yields with $m$ different maturities  $\mathbf{y}_{t}\in\mathbb{R}^{m}$ is of order 1, and is given by \citep{luetkepohl:2007}
\begin{equation}
\mathbf{y}_{t}=\boldsymbol{\alpha}+\mathbf{B}\mathbf{y}_{t-1}+\boldsymbol{\varepsilon}_{t},\qquad\boldsymbol{\varepsilon}_{t}\sim\text{ iid }N(\mathbf{0}_{m},\mathbf{P}^{-1}),\qquad t=2,\dots,n,\label{eq:VAR-model}
\end{equation}
where  $\boldsymbol{\alpha}\in\mathbb{R}^{m}$ and $\mathbf{B}\in\mathbb{R}^{m\times m}$ contain the VAR coefficients and $\mathbf{P}\in\mathbb{R}^{m\times m}$ is the positive definite precision  matrix of the Gaussian innovations $\boldsymbol{\varepsilon}_{t}$.
For the precision matrix, we use the parametrization $\mathbf{P}=\mathcal{P}(\mathbf{z})$ as defined in Equation (\ref{eq:MTP2-model-reparametrized}).
The spectral radius of $\mathbf{B}$ is
$\text{\ensuremath{\rho}(\ensuremath{\mathbf{B}})}=\max_{i}|\lambda_{i}|$,
where $\lambda_{i},\;i=1,\dots,m$ are the (possibly complex) eigenvalues
of $\mathbf{B}$, and the VAR model is stationary if \citep{luetkepohl:2007}
\begin{equation}
\rho(\mathbf{B})<1.\label{eq:VAR-stationary}
\end{equation}

To analyse the dynamic interaction between variables  in a VAR system, it is common to examine the response of one variable to an orthogonal impulse in another variable  of the system.
For the VAR model (\ref{eq:VAR-model}), a sequence of such responses of variable  $y_{i,t}$ over time  to  an impulse in variable $y_{j,t}$ in period $t$ (known as the impulse-response function, IRF) is given by \citep{luetkepohl:2007}
\begin{equation}
\frac{\partial y_{i,t+\ell}}{\partial u_{j,t}}= [\mathbf{B}^{\ell} \mathbf{G}]_{i,j},\qquad u_{j,t}=  [\mathbf{G}^{-1}\boldsymbol{\varepsilon}_{t}]_j, \qquad \ell=1,2,\ldots,\label{eq:VAR-IRF}
\end{equation}
where $\mathbf{G}$ is the unit lower-triangular matrix in the Cholesky decomposition of the covariance matrix $\mathbf{P}^{-1}$ of the VAR innovations $\boldsymbol{\varepsilon}_{t}$.

We apply the VAR model (\ref{eq:VAR-model}) with the stationarity constraint (\ref{eq:VAR-stationary}) to $m=6$ historical monthly interest rates on US zero-coupon bonds with maturities of 3, 6, 12, 24, 60 and 120 months from January 1983 to September 2003. The data are a subset of the data set analysed by \citet{https://doi.org/10.1002/jae.1220} (see also \citealp{CARRIERO20122026}). We set independent Gaussian priors for the VAR coefficients, $\boldsymbol{\alpha}_{i}\sim N(0,4^{2})$ and
$\mathbf{B}_{i,j}\sim N(0,4^{2})$, where the Gaussian prior for $\mathbf{B}$ is combined with the stationarity constraint. For $\mathbf{P}$ we assume a Wishart prior  with scaling matrix $\mathbf{I}_m$ and $m+10$ degrees of freedom.

To improve numerical efficiency of the sampling from the posterior we implement the (constrained) NGRHMC procedure as described in Sections 2 and 3 not directly for the VAR coefficient $(\boldsymbol{\alpha}, \mathbf{B})$  but the  OLS standardized versions thereof denoted by $(\tilde{\boldsymbol{\alpha}}, \tilde{\mathbf{B}})$. Thus the NGRHMC target density as introduced in Section 2 is $\pi(\mathbf{q})$ for $\mathbf{q}=({\tilde{\boldsymbol{\alpha}}}^T, \text{vec}(\tilde{\mathbf{B}})^T,\mathbf{z}^{T})^T$. The standardized VAR coefficients are defined by the linear transformations $(\boldsymbol{\alpha}_{i},\mathbf{B}_{i,\cdot})^{T}=\boldsymbol{\mu}^{(i)}+\boldsymbol{\Psi}^{(i)}  (\tilde{\boldsymbol{\alpha}}_{i},\tilde{\mathbf{B}}_{i,\cdot})^T  ,\;i=1,\dots,m$, where $\boldsymbol{\mu}^{(i)}$ is the vector of the OLS estimates for the coefficients in the $i$th linear regression of the VAR model $(\boldsymbol{\alpha}_{i},\mathbf{B}_{i,\cdot})^{T}$, and $\boldsymbol{\Psi}^{(i)}$ the lower triangular Cholesky factor of the estimated covariance matrix of the OLS estimator for $(\boldsymbol{\alpha}_{i},\mathbf{B}_{i,\cdot})^{T}$. The constrained NGRHMC is implemented as set out in Section 4.4, where the specific form of the constraint is $F(\text{vec}(\mathbf{B}))=1-\rho(\mathbf{B})\geq 0$ for the (non-trivial) affine relation between $\text{vec}(\mathbf{B})$ and $\mathbf{q}$ of the form $\text{vec}(\mathbf{B})=\mathbf{A}\mathbf{q}+\text{\ensuremath{\mathbf{b}}}$.

\begin{figure}
\begin{centering}
\includegraphics[scale=0.5]{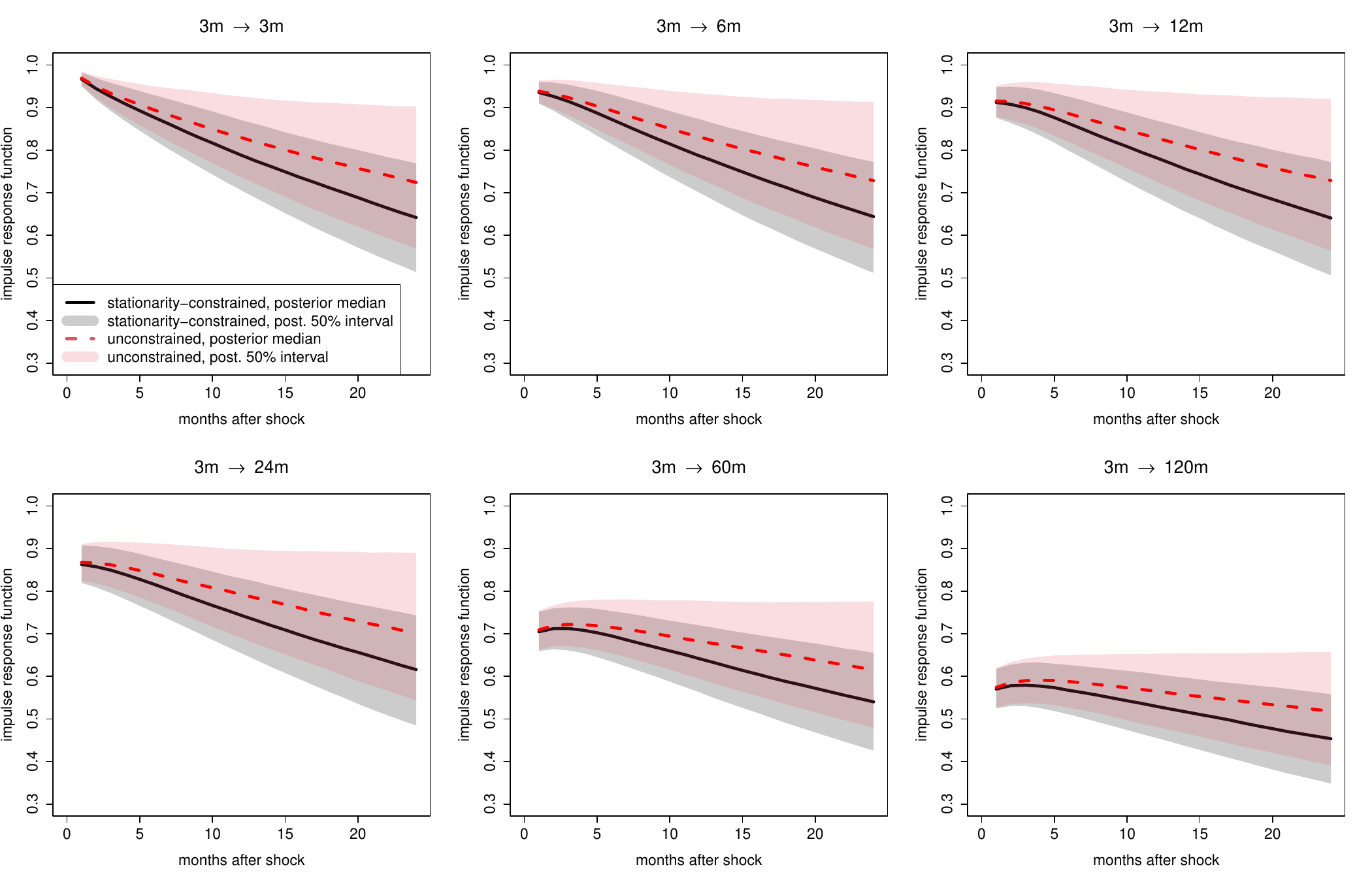}\caption{\label{fig:VAR IRF} Estimated posterior medians and 50\% credible intervals of the impulse responses of the bond yields $\mathbf{y}_{t}$ with maturities of 3, 6, 12, 24, 60 and 120 months to a 1-standard-deviation shock to the bond yield with
a maturity of 3 months for the VAR model (\ref{eq:VAR-model})  with and without the stationarity constraint (\ref{eq:VAR-stationary}).}
\par\end{centering}
\end{figure}

\begin{figure}
\begin{centering}
\includegraphics[scale=0.7]{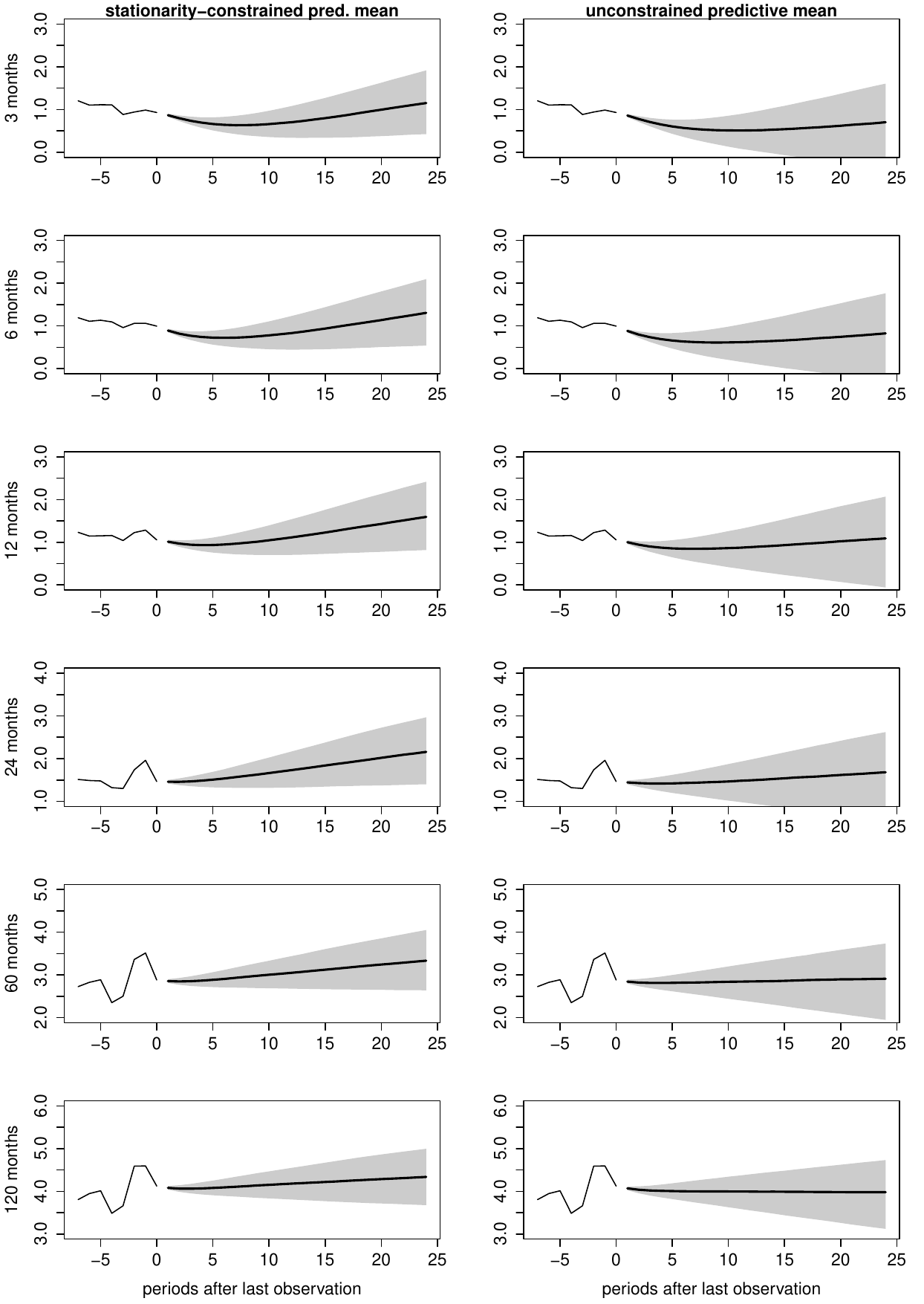}\caption{\label{fig:Predictive-means-1-24}
Estimated medians and 50\% intervals of the predictive density  $\pi(\mathbf{y}_{T+h}|\mathbf{y}_{1},\ldots,\mathbf{y}_{n})$ for forecasting the bond yields $\mathbf{y}_{t}$ for horizions $h=1,2,\ldots,24$ months ahead, resulting  for the VAR model (\ref{eq:VAR-model})  with (left panels) and without  the stationarity constraint (\ref{eq:VAR-stationary}) (right panels).  Thick black line is the estimated predictive
median, and the shaded region indicates 50\% intervals.  Thin line represent the last 8 observations of the data set.}
\par\end{centering}
\end{figure}

The diagnostic results for the simulation efficiency of the NGRHMC for the VAR parameters with and without the stationarity constraint are summarized in Table \ref{tab:Diagnostics-information-for-VAR}, and the  histograms and trace plots for the resulting samples from the posterior of the
spectral radius $\rho(\mathbf{B})$  are shown in Figure \ref{fig:VAR_rho_histogram}.
As can be seen in Figure \ref{fig:VAR_rho_histogram}, the proportion of the posterior probability mass in the non-stationary region resulting for the unconstrained VAR is quite large. The resulting estimate of the associated  posterior probability $P\{\rho(\mathbf{B})\geq1|\mathbf{y}_1,\ldots, \mathbf{y}_n\}$ is 0.31.
The ESS/s values  in Table \ref{tab:Diagnostics-information-for-VAR} show that the sampling efficiency of the constrained NGRHMC is moderately lower  than that of the  NGRHMC  for the unconstrained posterior.
Another benchmark for the sampling efficiency of the constrained NGRHMC is the efficiency of the brute force approach to sampling from the constrained posterior, which involves drawing from the unconstrained posterior  and discarding the parameter draws that violate the constraint.
Its ESS/s values can be approximated by multiplying the ESS/s values for the unconstrained NGRHMC in Table \ref{tab:Diagnostics-information-for-VAR} by the estimate for  $P\{\rho(\mathbf{B})< 1|\mathbf{y}_1,\ldots, \mathbf{y}_n\}$ found for the unconstrained posterior.
The resulting minimum for the ESS/s across all VAR parameters is 2.1, showing that the constrained NGRHMC (with a minimum ESS/s of 2.6) performs slightly better compared to the brute force approach.
Of course the performance of the brute force method will deteriorate as the posterior probability $P\{\rho(\mathbf{B})\geq1|\mathbf{y}_1,\ldots, \mathbf{y}_n\}$ increases, and thus it can be expected that this method has a very low simulation efficiency
in applications where this probability is much larger than the 0.31 found in our application.

Figure \ref{fig:VAR IRF} compares the posterior of the impulse responses for the VAR with and without the stationarity restriction. It displays the estimated posterior medians  together with the 50\% credible intervals of the impulse responses of all 6 bond yields  to a 1-standard-deviation shock to the bond yield with the shortest  maturity. It can be seen  that the estimated medians of the impulse responses for the constrained VAR are uniformly smaller than for the unconstrained VAR. This can be explained by the fact that the stationarity restriction moves the posterior of the autoregressive matrix $\mathbf{B}$ into a region that is associated with  a lower persistence  of the VAR innovations.  The result that the credible intervals of the impulse responses   for the constrained VAR are systematically tighter than for the unconstrained VAR reflects the regularizing effect of the constraint on the posterior of $\mathbf{B}$. Figure \ref{fig:Predictive-means-1-24} illustrates the effects of the stationary constraint on  out-of-sample forecasts of the bond yields based on the predictive density $\pi(\mathbf{y}_{T+h}|\mathbf{y}_{1},\ldots,\mathbf{y}_{n})$ for the Bayesian VAR model. This figure plots the estimated medians and 50\% intervals of the predictive density for forecast horizons of $h=1,2,\ldots,24$ months. The results show that the predictive medians for the stationarity-constrained VAR differ only slightly from their unconstrained counterparts.
However, we see larger differences in the prediction intervals, which are significantly smaller for the stationarity-constrained VAR than for the unconstrained one. This decrease in the forecast uncertainty indicates that the regularizing effect of the constraint on the parameter estimates also translates into the forecasts.

\section{Discussion}
We have proposed constrained numerical generalized randomized Hamiltonian Monte Carlo (NGRHMC) processes for numerically efficiently exploring distributions for constrained data or parameter spaces. 
They combine unconstrained NGRMHC processes for the interior of the constrained space with a randomized and sparse transition  kernel to properly update the Hamiltonian flow at the boundary of the constrained space so that it remains in that space. 
Our approach thus exploits the capability  of NGRMHC processes to efficiently explore complex distributions even in high-dimensional applications and combines this with a boundary transition kernel that, through its randomness and sparsity, produces updates to the Hamiltonian flow that efficiently explore the constrained space near its boundary.  The proposed approach thus provides a flexible and generic tool for numerically accurate approximation of complex distributions with various types of constraints, even in high-dimensional applications. 
It can therefore advance the development and use of Bayesian models with useful constrained priors that are difficult to handle with existing methods. 
The attractive performance of the proposed constrained GRHMC approach has been established both by simulation experiments and by posterior analyses of several Bayesian models for real-world datasets with challenging parameter space constraints.

While we have focused in this paper on simulating target distributions with a truncated domain, the numerical efficiency of the proposed NGRHMC approach motivates ongoing research in which we consider NGRHMC processes for the exploration of continuous target distributions with discontinuous gradients. These could then be used to simulate censored distributions, i.e. mixtures of a continuous distribution and a delta point mass. We also  propose to investigate the use of the NGRHMC approach  to handle discontinuous  target distributions  \citep{10.1093/biomet/asz083}.

\appendix
\renewcommand{\theequation}{A-\arabic{equation}} 
\setcounter{equation}{0}  
\renewcommand{\thesection}{A\arabic{section}} 
\setcounter{section}{0}  
\renewcommand{\thefigure}{A-\arabic{figure}} 
\setcounter{figure}{0}  
\renewcommand{\thetable}{A-\arabic{table}} 
\setcounter{table}{0}  

\section{\label{subsec:Correctness-of-()}Appendix: Correctness of the constrained GRHMC with the randomized transition kernel (\ref{eq:randomized_boundary_kernel})}
To show the correctness of the constrained GRHMC with the randomized boundary collision  transition kernel defined in Equation (\ref{eq:randomized_boundary_kernel}), it is sufficient to prove that this kernel satisfies the conditions given in Equations (\ref{eq:general_preservation_std_normal}) and (\ref{eq:IP_reversal}) \citep{BIERKENS2018148}.

Condition (\ref{eq:IP_reversal}) is satisfied since the inner product of the normal vector $\bar{\mathbf{n}}_{r}(\bar{\mathbf{q}})$ and the momentum update vector $\bar{\mathbf{p}}$ generated by the randomized boundary collision kernel is (suppressing the dependence of $\bar{\mathbf{n}}_{r}(\bar{\mathbf{q}})$ on $\bar{\mathbf{q}}$  in  the notation for the sake of simplicity)
\begin{align}
\bar{\mathbf{p}}^{T}\bar{\mathbf{n}}_{r}= & \underbrace{\left[\mathbf{z}-\frac{\mathbf{z}{}^{T}\bar{\mathbf{n}}_{r}}{\bar{\mathbf{n}}_{r}^{T}\bar{\mathbf{n}}_{r}}\bar{\mathbf{n}}_{r}\right]^{T}\bar{\mathbf{n}}_{r}}_{=0}-\frac{\left(\bar{\mathbf{p}}^{\prime}\right){}^{T}\bar{\mathbf{n}}_{r}}{\bar{\mathbf{n}}_{r}^{T}\bar{\mathbf{n}}_{r}}\bar{\mathbf{n}}_{r}^{T}\bar{\mathbf{n}}_{r}\\
= & -\left(\bar{\mathbf{p}}^{\prime}\right){}^{T}\bar{\mathbf{n}}_{r}, \nonumber
\end{align}
for any value of $\mathbf{z}$.

To show that condition (\ref{eq:general_preservation_std_normal}) is satisfied, we rewrite the momentum update vector $\bar{\mathbf{p}}$ generated by the randomized boundary collision kernel (\ref{eq:randomized_boundary_kernel}) as
\begin{align}\label{eq_app:momentum_update}
\bar{\mathbf{p}}=\mathbf{H}_{\bar{\mathbf{q}}}\left[\begin{array}{c}
\bar{\mathbf{p}}^{\prime}\\
\mathbf{z}
\end{array}\right],\quad\mbox{with}\quad \mathbf{H}_{\bar{\mathbf{q}}}=\left[\begin{array}{cc}
-\mathbf{R}_{\bar{\mathbf{q}}} & (\mathbf{I}_{d}-\mathbf{R}_{\bar{\mathbf{q}}})\end{array}\right],\quad \mathbf{z}\sim N(\mathbf{0}_{d},\mathbf{I}_{d}),
\end{align}
where the matrix $\mathbf{R}_{\bar{\mathbf{q}}}$ is defined in Equation (\ref{eq:collition_transition_kernel}), repeated here for convenience: $\mathbf{R}_{\bar{\mathbf{q}}}=\left(\bar{\mathbf{n}}_{r}^{T}\bar{\mathbf{n}}_{r}\right)^{-1}\bar{\mathbf{n}}_{r}\bar{\mathbf{n}}_{r}^{T}$.
Because $\mathbf{R}_{\bar{\mathbf{q}}}$ is a projection matrix with the properties $\mathbf{R}_{\bar{\mathbf{q}}}=\mathbf{R}_{\bar{\mathbf{q}}}^T$ and $\mathbf{R}_{\bar{\mathbf{q}}} \mathbf{R}_{\bar{\mathbf{q}}}^T=\mathbf{R}_{\bar{\mathbf{q}}}$,  it holds that
\begin{align}\label{eq_app:coefficient_matrix_H}
\mathbf{H}_{\bar{\mathbf{q}}}\mathbf{H}_{\bar{\mathbf{q}}}^{T}&=\mathbf{R}_{\bar{\mathbf{q}}} \mathbf{R}_{\bar{\mathbf{q}}}^T+(\mathbf{I}_{d} -\mathbf{R}_{\bar{\mathbf{q}}})(\mathbf{I}_{d}-\mathbf{R}_{\bar{\mathbf{q}}})^T\\
&=\mathbf{I}_{d}.\nonumber
\end{align}
From Equations (\ref{eq_app:momentum_update}) and (\ref{eq_app:coefficient_matrix_H}) it follows that if $\bar{\mathbf{p}}^{\prime}\sim N(\mathbf{0}_{d},\mathbf{I}_{d})$ such that $\left[(\bar{\mathbf{p}}^{\prime})^{T}\;\mathbf{z}^{T}\right]^{T}\sim N(\mathbf{0}_{2d},\mathbf{I}_{2d})$, then $\bar{\mathbf{p}}\sim N(\mathbf{0}_{d},\mathbf{I}_{d})$. Thus, the randomized boundary collision kernel leaves the standard normal distribution of the momentum invariant, as stated in condition (\ref{eq:general_preservation_std_normal}).

\pagebreak

\bibliographystyle{chicago}
\bibliography{kleppe}

\end{document}